\pdfoutput=1

\documentclass[11pt]{article}

\usepackage[final]{acl}
\usepackage{fc_custom}

\usepackage{times}
\usepackage{latexsym}
\usepackage{multirow}

\usepackage[T1]{fontenc}

\usepackage[utf8]{inputenc}
\usepackage{xspace}    

\usepackage{microtype}
\usepackage{booktabs}
\usepackage{graphicx}
\usepackage{diagbox}
\usepackage{longtable}
\usepackage{tabularx}
\usepackage{tabularray}
\usepackage{adjustbox}
\usepackage{amsmath}
\usepackage{amssymb}

\usepackage{bbm}
\usepackage{bbold}

\usepackage{amssymb}
\usepackage{algorithm}
\usepackage{algorithmicx}
\usepackage{algpseudocode}
\usepackage{todonotes}
\usepackage{siunitx}

\title{\shorttitle: Enhancing Retriever Generalization for Scientific Domain \\ through Complementary Granularity}

\newcommand{\tong}[1]{
\ifthenelse{\equal{\commentisunseen}{0}}{
{{\textcolor{purple}{#1}}}}
{}\xspace
}

\newcommand{\sihao}[1]{
\ifthenelse{\equal{\commentisunseen}{0}}{
{\textcolor{orange}{[Sihao: #1]}}
{}\xspace
}
}
\newcommand{\xinran}[1]{
\ifthenelse{\equal{\commentisunseen}{0}}{
{{\textcolor{red}{#1}}}}
{}\xspace
}

\author{
\textbf{Fengyu Cai}\textsuperscript{1} \,
\textbf{Xinran Zhao}\textsuperscript{2} \,
\textbf{Tong Chen}\textsuperscript{3} \,
\textbf{Sihao Chen}\textsuperscript{4} \, \\
\textbf{Hongming Zhang}\textsuperscript{5} \,
\textbf{Iryna Gurevych}\textsuperscript{1} \,
\textbf{Heinz Koeppl}\textsuperscript{1}
\vspace{5pt} \\ 
\textsuperscript{1}Technical University of Darmstadt \,
\textsuperscript{2}Carnegie Mellon University \,
\textsuperscript{3}University of Washington \, \\
\textsuperscript{4}University of Pennsylvania \,
\textsuperscript{5}Tencent AI Lab \\
\texttt{\{fengyu.cai, heinz.koeppl\}@tu-darmstadt.de}
}

\begin{document}
\maketitle

\begin{abstract}
Recent studies show the growing significance of document retrieval in the generation of LLMs, i.e., RAG, within the scientific domain by bridging their knowledge gap.
However, dense retrievers often struggle with domain-specific retrieval and complex query-document relationships, particularly when query segments correspond to various parts of a document.
To alleviate such prevalent challenges, this paper introduces \shorttitle, which improves dense retrievers' awareness of query-document matching across various levels of granularity in queries and documents using a zero-shot approach.
\shorttitle fuses various metrics based on these granularities to a united score that reflects a comprehensive query-document similarity.
Our experiments demonstrate that \shorttitle outperforms previous document retrieval by 24.7\% and 9.8\% on nDCG@5 with unsupervised and supervised retrievers, respectively, averaged on queries containing multiple subqueries from five scientific retrieval datasets.
Moreover, the efficacy of two downstream scientific question-answering tasks highlights the advantage of \shorttitle to boost the application of LLMs in the scientific domain. The code and experimental datasets are available. \footnote{\url{https://github.com/TRUMANCFY/MixGR}}

\end{abstract}

\section{Introduction}

\begin{figure}[t]
    \centering
    \begin{subfigure}[b]{0.5\textwidth}  %
        \centering
        \includegraphics[width=\textwidth]{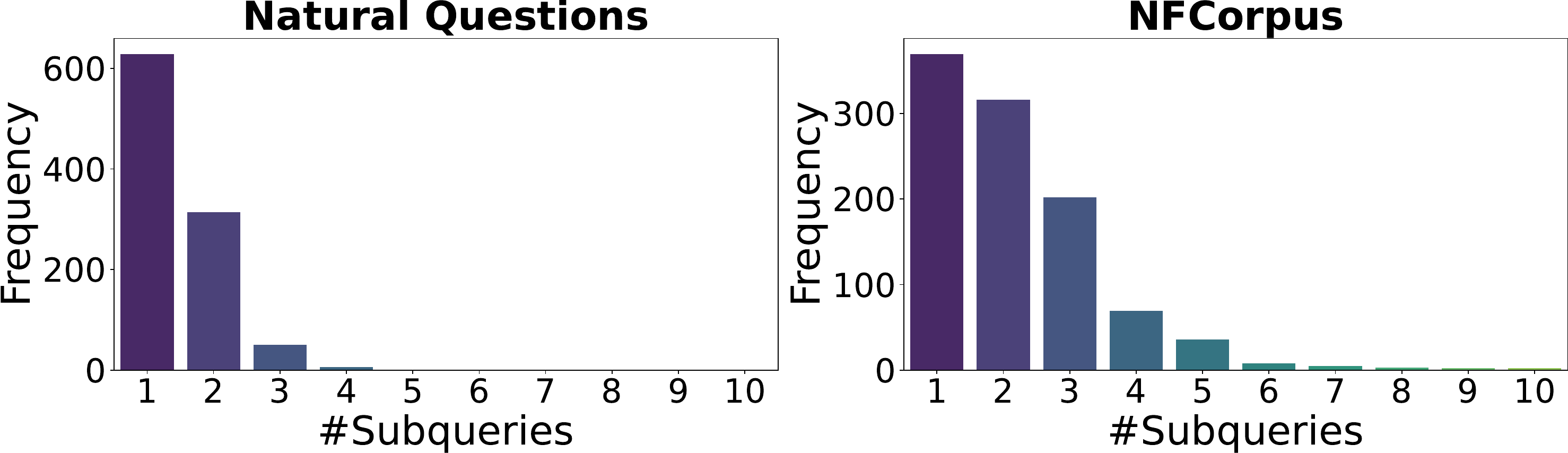}  %
        \caption{Subquery distribution of general and scientific queries: scientific queries, e.g., NFCorpus (\citealt{boteva2016full}, \ti{Right}), demonstrate a more diverse range of subqueries per query than general queries, e.g., Natural Questions (\citealt{kwiatkowski2019natural}, \ti{Left}).}
        \label{fig:subquery-dist}
    \end{subfigure}
    \hfill
    \begin{subfigure}[t]{0.5\textwidth} %
        \includegraphics[width=\textwidth]{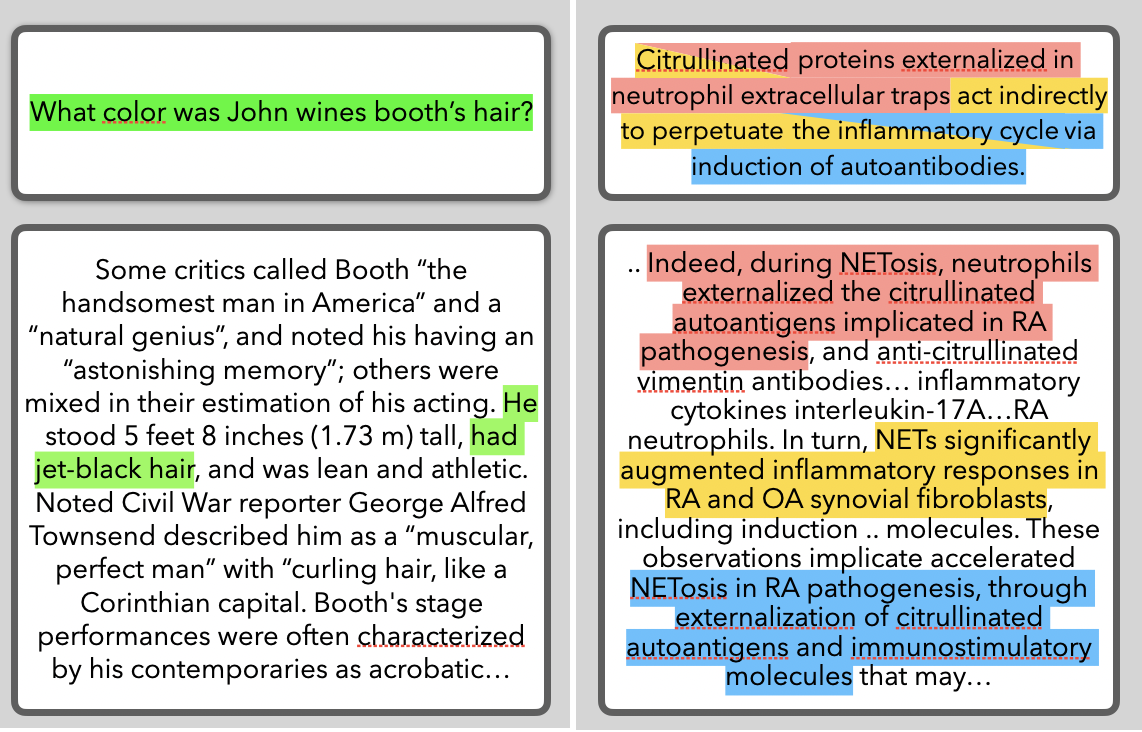} %
        \caption{Comparison between general and scientific query-doc retrieval: compared with the general query-doc retrieval exemplified by NQ (\citealt{kwiatkowski2019natural}, \ti{Left}), the scientific query-doc retrieval exemplified by SciFact (\citealt{wadden-etal-2020-fact}, \ti{Right}) demonstrates that one query can be decomposed to multiple subqueries, which can be mapped to different parts of documents.}
        \label{fig:query-doc}
    \end{subfigure}
    \caption{Scientific document retrieval is shown to be more complicated than general domains.}
    \label{fig:both_figures}
\end{figure}

Recent advances in Large Language Models (LLMs) have significantly impacted various scientific domains \cite{DBLP:journals/corr/abs-2205-01068,DBLP:journals/corr/abs-2307-09288,birhane2023science,grossmann2023ai}.
However, LLMs are notorious for their tendency to produce hallucinations, generating unreliable outputs \cite{DBLP:journals/csur/JiLFYSXIBMF23}.
To address this, Retrieval-Augmented Generation (RAG; \citealt{lewis2020retrieval}) has been developed to address this issue by incorporating external knowledge during the generation.

\begin{figure*}[t]
    \centering
    \includegraphics[width=\textwidth]{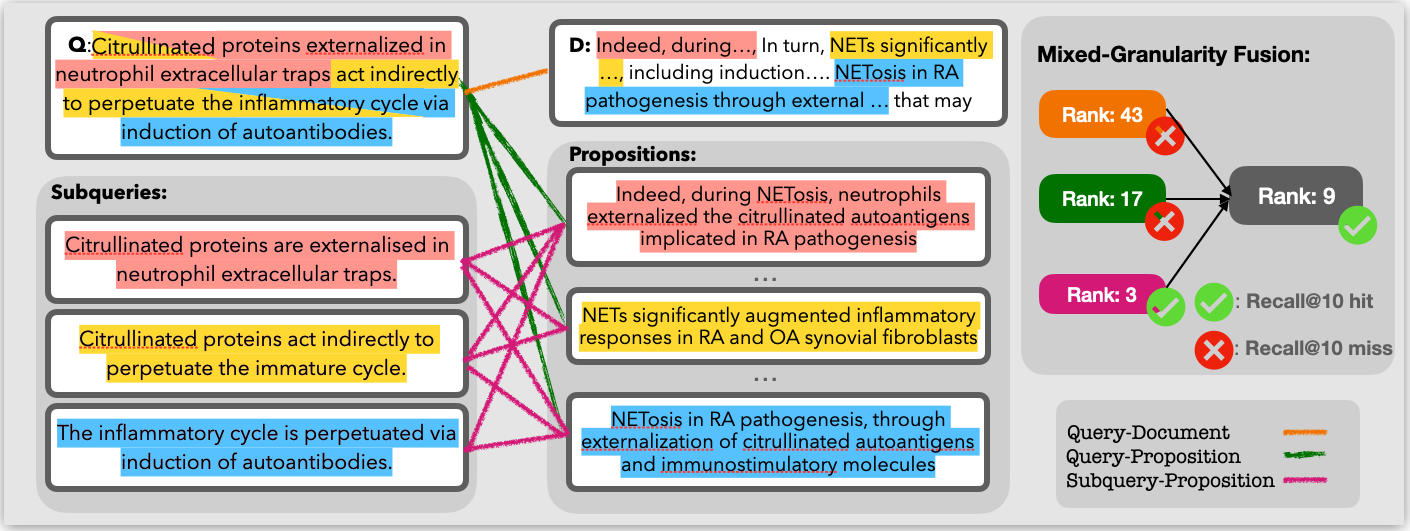} %
    \caption{The illustration of \shorttitle: Both queries and documents (e.g., the query-doc pair from SciFact in Figure \ref{fig:query-doc}) are decomposed into subqueries and propositions, respectively, each containing distinct semantic components. Starting from the original queries and documents along with their decomposed elements, metrics from various granularity combinations are fused into a single integrated score.
    }
    \label{fig:rrf}
\end{figure*}

Though notable for accessing external and relevant knowledge, dense retrievers face specific challenges in the scientific domain:
(1) \ti{Domain-specific} nature:
dense retrievers are typically trained on the general corpus such as Natural Questions (NQ; \citealt{kwiatkowski2019natural}).
However, scientific domains differ notably, e.g., the terminology and the pattern of queries as shown in Figure \ref{fig:subquery-dist}.
(2) \ti{Complexity} of scientific documents: they are long, structured \cite{erera-etal-2019-summarization} and contain complex relationships between arguments \cite{stab2014argumentation}.
Figure \ref{fig:subquery-dist} demonstrates that scientific queries tend to contain more subqueries than those in general domains.
This indicates that subqueries within a single query may align with different parts of a document (doc), resulting in complex interactions between queries and documents (Figure \ref{fig:query-doc}).
Such complexity poses significant challenges for dense retrievers \cite{lupart2023ms}.
Addressing these challenges requires specific training on the scientific corpus.
However, this is often hindered by the necessity of extensive annotations \cite{wadden-etal-2020-fact} and extra computation \cite{wang-etal-2021-tsdae-using}.

In this study, we introduce a novel \tf{zero-shot} approach that effectively adapts dense retrievers to scientific domains.
This method specifically addresses the complexities arising from the composition of scientific queries and their consequent intricate relationships with documents.
Inspired by \citet{chen2023dense}, showing that finer units improve retrievers' generalization to rare entities, we incorporate more granular retrieval units, specifically propositions (prop), to address domain-specific challenges as shown in Figure \ref{fig:rrf}.
Given the complexity between scientific queries and documents (Figure \ref{fig:query-doc}), we also consider finer units within queries--subqueries--to measure query-doc similarity at a finer granularity.
This metric captures the similarity between subqueries and propositions, moving beyond simple point similarity between query-doc vectors.
Given a query, the distribution of corresponding information within a document is unknown.
Additionally, our empirical analysis reveals that similarities at various granularities provide complementary insights. 
Therefore, for each query-doc pair, we fuse the metrics from these granularities to a unified score, termed \tf{Mix}ed-\tf{G}ranularity \tf{R}etrieval as \shorttitle, as depicted in Figure \ref{fig:rrf}.

We conducted document retrieval experiments on five scientific datasets using six dense retrievers, comprising two unsupervised and four supervised models.
Our results demonstrate that \shorttitle markedly surpasses previous query-doc retrieval methods.
Notably, we recorded an average improvement of 24.7\% for unsupervised retrievers and 9.8\% for supervised retrievers in terms of nDCG@5 for queries involving multiple subqueries.
Furthermore, documents retrieved via \shorttitle substantially enhance the performance of downstream scientific QA tasks, underscoring their potential utility for RAG within scientific domains.

Our contributions are three-fold:
\begin{itemize}[leftmargin=*, itemsep=0em]
    \item We identify the challenges within scientific document retrieval, i.e., domain shift and query-doc complexity.
    We initiate retrieval with mixed granularity within queries and documents to address these issues;
    \item We propose \shorttitle, which further incorporates finer granularities within queries and documents, computes query-doc similarity over various granularity combinations and fuses them as a united score.
    Our experiments across five datasets and six dense retrievers empirically reveal that \shorttitle significantly enhances existing retrievers on the scientific document retrieval and downstream QA tasks;
    \item Further analysis demonstrates the complementarity of metrics based on different granularities and the generalization of \shorttitle in retrieving units finer than documents.
\end{itemize}

\section{Preliminary and Related works}
\paragraph{Generalization of Dense Retrievers}
Dense retrievers generally employ a dual-encoder framework \cite{yih-etal-2011-learning,reimers-gurevych-2019-sentence} to separately encode queries and documents into compact vectors and measure relevance using a non-parametric similarity function \cite{mussmann2016learning}.
However, the simplicity of the similarity function (e.g., cosine similarity) can restrict expressiveness, leading to suboptimal generalization in new domains such as scientific fields that differ from original training data \cite{thakur2021beir}.
To improve dense retrievers' adaptability across tasks, researchers have used data augmentation \cite{wang-etal-2022-gpl,lin-etal-2023-train,dai2023promptagator}, continual learning \cite{chang2020pre,sachan-etal-2021-end,oguz-etal-2022-domain}, and task-aware training \cite{xin-etal-2022-zero,cheng-etal-2023-task}. 
However, these methods still require training on domain-specific data, incurring additional computational costs. This work focuses on \textit{zero-shot} generalization of dense retrievers to scientific fields by incorporating multi-granularity similarities within queries and documents.

\paragraph{Granularity in Retrieval}
For dense retrieval, the selection of the retrieval unit needs to balance the trade-off between completeness and compactness.
Coarser units, like documents or fixed-length passages, theoretically encompass more context but may introduce extraneous information, adversely affecting retrievers and downstream tasks \cite{shi2023large,wang2023learning}.
Conversely, finer units like sentences are not always self-contained and may lose context, thereby hindering retrieval \cite{akkalyoncu-yilmaz-etal-2019-cross,yang2020multilingual}.
Additionally, some studies extend beyond complete sentences; for example, \citet{lee-etal-2021-learning-dense} use phrases as learning units to develop corresponding representations. 
Meanwhile, ColBERT \cite{khattab2020colbert} addresses token-level query-doc interaction but is hampered by low efficiency.

\citet{chen2023dense} propose using \ti{propositions} as retrieval units, defined as atomic expressions of meaning \cite{min-etal-2023-factscore}.
These units are contextualized and self-contained, including necessary context through decontextualization, e.g., coreference resolution \cite{zhang-etal-2021-brief}.
Proposition retrieval improves retrieval of documents with long-tail information, potentially benefiting domain-specific tasks.
This motivates the use of propositions as retrieval units for scientific document retrieval.
Furthermore, we extend fine granularity to queries and enhance the query-doc similarity measurement, moving from a point-wise assessment between two vectors to integrating multiple query-doc granularity combinations.

\paragraph{Fusion within Retrieval}
Each type of retriever, sparse or dense, has its own strength and can be complementary with each other.
Based on this insight, previous studies have explored the fusion of searches conducted by different retrievers as a zero-shot solution for domain adaptation \cite{thakur2021beir}.
A common method involves the convex combination, which linearly combines similarity scores \cite{karpukhin-etal-2020-dense,wang2021bert,ma2021replication}.
However, this approach is sensitive to the weighting of different metrics and score normalization, which complicates configuration across different setups \cite{chen2022out}.

In this work, we enhance retrieval by integrating searches across various query and document granularity levels for a given retriever.
To avoid the limitations of convex combination on parameter searching, we use Reciprocal Rank Fusion (RRF; \citealt{cormack2009reciprocal}), a robust, non-parametric method \cite{chen2022out}, to aggregate these searches.

\section{\shorttitle: Mix-Granularity Retrieval} \label{sec:method}

\subsection{Finer Units in Queries and Documents} \label{sec:query-document}
We first decompose queries and documents into atomic units, i.e., subqueries and propositions, respectively.
A proposition (or subquery) should meet the following three principal criteria \cite{min-etal-2023-factscore}:

\begin{itemize}[leftmargin=*, itemsep=0em]
    \item Each proposition conveys a distinct semantic unit, collectively expressing the complete meaning.
    \item Propositions should be atomic and indivisible.
    \item According to \citet{choi2021decontextualization}, propositions should be contextualized and self-contained, including all necessary text information such as resolved coreferences for clear interpretation.
\end{itemize}

Here, we employ an off-the-shelf model, \ti{propositioner},\footnote{{\url{https://huggingface.co/chentong00/propositionizer-wiki-flan-t5-large}}} for decomposing queries and documents \cite{chen2023dense}.
This model is developed by distilling the decomposition capacities of GPT-4 \cite{achiam2023gpt} to a Flan-T5-Large model \cite{chung2024scaling} using Wikipedia as the corpus.
We sample decomposition results from 100 queries and 100 documents from the datasets in \S \ref{subsec:datasets} and manually label the correctness of decomposition as shown in Table \ref{tab:prop-annotation}.
This model is shown to effectively decompose queries and documents into atomic units within scientific domains.
Please see Appendix \ref{appendix:examples} for further details.

\newcommand{\plusminus}{\mathbin{\mathpalette\pmop\relax}}
\newcommand{\pmop}[2]{\ooalign{%
  \raisebox{.1\height}{$#1+$}\cr
  \smash{\raisebox{-.6\height}{$#1-$}}\cr}}
  
\begin{table}[t]
    \centering
    \resizebox{0.3\textwidth}{!}{ %
        \begin{tabular}{lcc}
            \toprule
            & Query & Document \\
            \midrule
            Accuracy (\%) & 96.3  & 94.7 \\
            IAA (\%) & 92.0 & 89.0 \\
            \bottomrule
        \end{tabular}
    }
    \caption{Human-evaluated accuracy of query/document decomposition by \ti{propositioner} \cite{chen2023dense}.}
    \label{tab:prop-annotation}
    \vspace{-0.5em}
\end{table}

\subsection{Multi-Granularity Similarity Calculation}
\label{sec:method:multi-granularity}
Given these various granularities including queries, subqueries, documents, and propositions, we extend the query-doc similarity metrics to include measurements across different combinations of granularities as depicted in Figure \ref{fig:rrf}.

\paragraph{Notations}
The sets of queries and documents are denoted as $\mathcal{Q}$ and $\mathcal{D}$, respectively.
Given a retriever $s$, the similarity between a query $\mathrm{q} \in \mathcal{Q}$ and a document $\mathrm{d} \in \mathcal{D}$ is denoted as $s(\mathrm{q}, \mathrm{d})$.
A document $\mathrm{d}$ can be decomposed to $N$ propositions, i.e., $\mathrm{d} = [d_1, ..., d_N]$.
And a query $q$ can be decomposed to $M$ subqueries, i.e., $\mathrm{q} = [q_1, ..., q_M]$.

\paragraph{Query-doc $\boldsymbol{s_{q\text{-}d}}$:} The direct and original similarity between $\mathrm{q}$ and $\mathrm{d}$ is $\boldsymbol{s_{q\text{-}d}}(\mathrm{q}, \mathrm{d}) \equiv s(\mathrm{q}, \mathrm{d})$.

\paragraph{Query-prop $\boldsymbol{s_{q\text{-}p}}$:}Recent works \cite{chen2023dense} determine query-doc similarity by calculating the maximum similarity between the \tf{query} and individual \tf{propositions} within the document \cite{lee-etal-2021-phrase,chen2023dense}.
The computation of this metric, denoted as $s_{q\text{-}p}$, is as follows:
\begin{align}
    s_{q\text{-}p}(\mathrm{q}, \mathrm{d}) = \max_{i=1,...,N} \{s(\mathrm{q}, d_i)\}.
    \label{eq:query-prop}
\end{align}

\paragraph{Subquery-prop $\boldsymbol{s_{s\text{-}p}}$:} Considering that different parts of a query may be captured by various propositions within a document shown in Figure \ref{fig:query-doc}, we further assess query-doc similarity by analyzing the relationships between \tf{subqueries} and individual \tf{propositions}.
The similarity between a query and a document can be defined as the average similarity across subqueries, calculated by identifying the maximum similarity between one subquery and each proposition, in analogy to MaxSim in ColBERT \cite{khattab2020colbert}.
This metric, represented by $s_{s\text{-}p}$, is calculated as:

\begin{align}
    s_{s\text{-}p}(\mathrm{q}, \mathrm{d}) = \frac{1}{M} \sum_{i=1}^{M} \max_{j=1,...,N}{ \{s(q_i, d_j)\} }.
    \label{eq:sub-prop}
\end{align}

\subsection{Reciprocal Rank Fusion}
We then use RRF to fuse these metrics across different query and document granularities:

\begin{align}
    s_f(\mathrm{q}, \mathrm{d}) &= \frac{1}{1 + r_{q\text{-}d}(\mathrm{q}, \mathrm{d})} + \frac{1}{1 + r_{q\text{-}p}(\mathrm{q}, \mathrm{d})} \nonumber \\
    &+ \frac{1}{1 + r_{s\text{-}p}(\mathrm{q}, \mathrm{d})},
    \label{eq:rrf}
\end{align}
where $r_{q\text{-}d}$, $r_{q\text{-}p}$, $r_{s\text{-}p}$ $\in \mathbb{R}_{\geq 0}$ signify the rank of the retrieve results by $s_{q\text{-}d}$, $s_{q\text{-}p}$, and $s_{s\text{-}p}$, respectively.
Technically, we retrieve the top-$k$ results $R^k_{q\text{-}d}$, $R^k_{q\text{-}p}$, and $R^k_{s\text{-}p}$ by $s_{q\text{-}d}$, $s_{q\text{-}p}$, and $s_{s\text{-}p}$, respectively, where $k$ is set 200 empirically.
When a query-doc pair ($\mathrm{q}'$, $\mathrm{d}'$) in one retrieval result does not exist in the other sets (e.g., $(\mathrm{q}', \mathrm{d}') \in R^k_{q\text{-}d}$ but $(\mathrm{q}', \mathrm{d}') \notin R^k_{q\text{-}p}$), we will calculate the missing similarity (e.g., $s_{q\text{-}p}(\mathrm{q}',\mathrm{d}')$) before aggregation.

\begin{table*}[h]
\centering
\resizebox{0.98\textwidth}{!}{%
\begin{tabular}{lc|cc|cc|cc|cc|cc|c
@{\hspace{-6pt}}
S[table-format=2.1]
c
@{\hspace{-8pt}}
S[table-format=2.1]}
\toprule

\multirow{2}{*}{\tf{Retriever}} & \multicolumn{1}{c|}{\multirow{2}{*}{\tf{Setup}}} & \multicolumn{2}{c|}{\tf{BioASQ}} & \multicolumn{2}{c|}{\tf{NFCorpus}} & \multicolumn{2}{c|}{\tf{SciDocs}} & \multicolumn{2}{c|}{\tf{SciFact}} & \multicolumn{2}{c|}{\tf{SciQ}} & \multicolumn{4}{c}{\tf{Avg.}} \\
 & & ND@5  & ND@20 & ND@5  & ND@20 & ND@5  & ND@20 & ND@5  & ND@20 & ND@5  & ND@20  & ND@5 &  & ND@20 & \\ 
\midrule
\multicolumn{16}{c}{Unsupervised Dense Retrievers} \\
\midrule
\multirow{4}{*}{\tf{SimCSE}}
& $s_{q\text{-}d}$ & 17.0 & 17.0& 16.2 & 13.3& 7.6 & 9.7& 27.1 & 31.2& 62.3 & 67.3& 26.0 & & 27.7 & \\
& $s_{q\text{-}p}$ & 28.4 & 28.2& 20.0 & 16.4& \ul{8.2} & \ul{11.1} & \ul{32.8} & \ul{37.2} & 75.6 & 78.5& 33.0 & & 34.3 & \\
& $s_{s\text{-}p}$ & \tf{31.1} & \ul{30.6} & \tf{22.8} & \tf{18.3} & 7.3 & 10.5& 32.7 & 36.9& \ul{80.9} & \ul{83.2} & \ul{35.0} & & \ul{35.9} &  \\
& \shorttitle      & \ul{30.7} & \tf{31.3} & \ul{22.3} & \ul{18.1} & \tf{9.1} & \tf{12.2}& \tf{34.8} & \tf{39.8}& \tf{84.0} & \tf{85.5} & \tf{36.2} & {\tiny{+39.2\%}} & \tf{37.4} & \tiny{+35.0\%} \\
\midrule
\multirow{4}{*}{\tf{Contriever}}
& $s_{q\text{-}d}$ & \ul{64.8} & \ul{68.3}& 42.2 & 34.9& 13.5 & 18.5& \ul{64.5} & 68.5& 67.2 & 70.0& 50.5 & & 52.0 &\\
& $s_{q\text{-}p}$ & 64.1 & 68.2& \ul{43.0} & \ul{35.5} & \ul{14.5} & \ul{19.4} & 64.0 & \ul{68.9} & 79.7 & 81.0& \ul{53.1} & & \ul{54.6} & \\
& $s_{s\text{-}p}$ & 63.7 & 68.0& 41.4 & 34.9& 13.5 & 18.3& 63.2 & 67.5& \ul{83.6} & \ul{84.6} & \ul{53.1} & & \ul{54.6} & \\
& \shorttitle & \tf{67.0} & \tf{71.7}& \tf{44.0} & \tf{37.1}& \tf{15.5} & \tf{20.7} & \tf{66.4} & \tf{71.0} & \tf{85.2} & \tf{86.7} & \tf{55.6} & \tiny{+10.1\%} & \tf{57.5} & \tiny(+10.6\%) \\
\midrule
\multicolumn{16}{c}{Supervised Dense Retrievers} \\
\midrule
\multirow{4}{*}{\tf{DPR}}
& $s_{q\text{-}d}$ & 39.1 & 39.1& 25.1 & 20.7& 7.3 & 10.4& 31.8 & 37.7& 60.6 & 64.1& 32.8 & & 34.4 &\\
& $s_{q\text{-}p}$ & \ul{43.1} & \ul{43.8} & 25.2 & 20.6& \ul{7.8} & \ul{10.6} & 36.1 & 40.5& 63.6 & 67.9& 35.2 & & 36.7 & \\
& $s_{s\text{-}p}$ & 41.1 & 41.7& \ul{26.5} & \ul{21.4} & 6.4 & 10.0& \ul{37.1} & \ul{41.3} & \ul{67.7} & \ul{70.7} & \ul{35.8} & & \ul{37.0} & \\
& \shorttitle & \tf{44.6} & \tf{45.8} & \tf{27.7} & \tf{22.9} & \tf{8.2} & \tf{11.5} & \tf{39.4} & \tf{43.6} & \tf{73.6} & \tf{76.1}& \tf{38.7} & \tiny{+18.0\%} & \tf{40.0} & \tiny{+14.3\%} \\
\midrule
\multirow{4}{*}{\tf{ANCE}}
& $s_{q\text{-}d}$ & 48.8 & 48.2& 29.9 & 24.4& \ul{9.3} & \ul{13.1} & 41.5 & 45.3& \ul{66.4} & \ul{69.1} & 39.2 & & 40.0 & \\
& $s_{q\text{-}p}$ & \ul{53.0} & \ul{53.7} & 29.4 & 24.0& 9.2 & 12.9& 43.3 & 46.4& 62.3 & 66.4& \ul{39.4} & & \ul{40.7} &\\
& $s_{s\text{-}p}$ & 49.0 & 49.8 & \ul{30.3} & \ul{24.5} & 7.5 & 11.9& \ul{43.5} & \ul{47.3} & 66.1 & \ul{69.1} & 39.3 & & 40.5 &\\
& \shorttitle & \tf{53.4} & \tf{54.7} & \tf{31.9} & \tf{25.9} & \tf{9.6} & \tf{14.1} & \tf{46.8} & \tf{49.9} & \tf{74.4} & \tf{76.8} & \tf{43.2} & \tiny{+10.2\%} & \tf{44.3} & \tiny{+10.4\%} \\
\midrule
\multirow{4}{*}{\tf{TAS-B}}
& $s_{q\text{-}d}$ & 68.8 & 70.1& 42.3 & 34.1& 13.8 & \ul{19.3} & 60.1 & \ul{65.6} & 84.8 & 86.3& 54.0 & & \ul{55.1} & \\
& $s_{q\text{-}p}$ & \ul{70.9} & \ul{72.3} & \ul{42.5} & \ul{34.4} & \tf{14.3} & 18.1& 60.7 & 64.4& \ul{85.6} & 86.3& \ul{54.8} & & \ul{55.1} & \\
& $s_{s\text{-}p}$ & 67.0 & 69.4& 40.9 & 33.1& 12.6 & 17.2& \ul{61.7} & 65.0& 85.3 & \ul{86.6} & 53.5 & & 54.3 & \\
& \shorttitle & \tf{71.7} & \tf{74.0} & \tf{43.6} & \tf{35.2} & \ul{14.0} & \tf{19.6} & \tf{62.7} & \tf{66.9} & \tf{90.5} & \tf{91.0} & \tf{56.5} & \tiny{+4.6\%} & \tf{57.3} & \tiny{+4.0\%} \\
\midrule
\multirow{4}{*}{\tf{GTR}}
& $s_{q\text{-}d}$ & 63.5 & 63.1& 42.1 & 34.1& \tf{13.6} & \ul{18.9} & 58.3 & 62.2& 83.3 & 84.4& 52.2 & & 52.5 &\\
& $s_{q\text{-}p}$ & \ul{65.9} & \ul{67.8} & \ul{42.3} & \ul{34.4} & \ul{13.2} & 18.0& \ul{60.6} & \ul{63.3} & 85.8 & 86.5& \ul{53.6} & & \ul{54.0} & \\
& $s_{s\text{-}p}$ & 61.2 & 63.5& 41.5 & 33.6& 11.6 & 16.2& 58.4 & 62.0& \ul{88.5} & \ul{89.0} & 52.3 & & 52.9 & \\
& \shorttitle & \tf{66.8} & \tf{68.6} & \tf{43.3} & \tf{35.6} & \tf{13.6} & \tf{19.2}& \tf{60.9} & \tf{64.5} & \tf{92.9} & \tf{93.0}& \tf{55.5} & \tiny{+6.3\%} & \tf{56.2} & \tiny{+7.0\%} \\
\bottomrule
\end{tabular}%
}
\caption{
Document Retrieval Performance (nDCG@$k$ = 5, 20 in percentage, abbreviated as ND@$k$): We evaluated five distinct scientific retrieval datasets using two unsupervised and four supervised retrievers.
The retrieval results were compared among various metrics: $s_{q\text{-}d}$ (previous query-doc similarity), $s_{q\text{-}p}$ \cite{chen2023dense}, $s_{s\text{-}p}$, and \shorttitle, as detailed in \S \ref{sec:method:multi-granularity}.
\tf{Bold} presents the best performance across the metrics, while \ul{underline} denotes the second-best performance.
\shorttitle outperforms all three other metrics, where the percentage in parentheses indicates the relative improvement compared with $s_{q\text{-}d}$.
}
\label{tab:ir}
\end{table*}

\section{Experimental Setting} \label{sec:experiments}
\subsection{Scientific Retrieval Datasets} \label{subsec:datasets}
We evaluate our approach on five different scientific retrieval tasks, including BioASQ \cite{tsatsaronis2015overview}, NFCorpus \cite{boteva2016full}, SciDocs \cite{cohan-etal-2020-specter}, SciFact \cite{wadden-etal-2020-fact}, and SciQ \cite{welbl-etal-2017-crowdsourcing}, as shown in Table \ref{tab:dataset_stats} in Appendix \ref{appendix:datasets}.
We employ the \ti{propositioner} released by \citet{chen2023dense} mentioned in 
\S\ref{sec:query-document} to break down both queries and documents into atomic units.
As we focus with priority on query-doc complexity in scientific domains, we report the experiments and analysis on the subset of the queries that contain multiple subqueries.
For queries containing only a single subquery, we also assess the methodology of \shorttitle by omitting $s_{s\text{-}p}$. The detailed results are shown in Table \ref{tab:ir_single} in Appendix \ref{appendix:single_subquery}.
The results encompassing all queries, ones containing both single and multiple subqueries, are detailed in Table \ref{tab:ir_single} located in Appendix \ref{appendix:single_subquery}.

\subsection{Dense Retrievers} \label{sec:dense-retriever}
We evaluate the performance of six off-the-shelf dense retrievers, both supervised and unsupervised.
Supervised retrievers are trained using human-labeled query-doc pairs in general domains,\footnote{The supervised retrievers used in our experiment have not been trained on these five datasets.} while unsupervised models do not require labeled data.
These retrievers encode the queries and index the corpus at both document and proposition levels:

\begin{itemize}[leftmargin=*, itemsep=0em]
    \item SimCSE \citep{gao-etal-2021-simcse} employs a BERT-base \cite{devlin-etal-2019-bert} encoder trained on randomly selected unlabeled Wikipedia sentences.
    \item Contriever \citep{izacard2022unsupervised} is an unsupervised retriever evolved from a BERT-base encoder, contrastively trained on segments from unlabelled web and Wikipedia documents.
    \item DPR \citep{karpukhin-etal-2020-dense} is built with a dual-encoder BERT-base architecture, finetuned on a suite of open-domain datasets with labels, such as SQuAD \cite{rajpurkar-etal-2016-squad}.
    \item ANCE \citep{xiong2021approximate} mirrors the configuration of DPR but incorporates a training scheme of Approximate Nearest Neighbor Negative Contrastive Estimation (ANCE).
    \item TAS-B \citep{hofstatter2021efficiently} is a dual-encoder BERT-base model distilled from ColBERT on MS MARCO \cite{nguyen2016ms}. %
    \item GTR \citep{ni-etal-2022-large} is a T5-base encoder, focusing on generalization, pre-trained on unlabeled QA pairs, and fine-tuned on labeled data including MS MARCO.
\end{itemize}

More details on retrievers and experimental setups are presented in Appendices \ref{appendix:retrievers} and \ref{appendix:indexing}.

\subsection{Document Retrieval Evaluation}
We assess the performance of \shorttitle in the task of document retrieval.
Due to input length limitations for retrievers \cite{karpukhin-etal-2020-dense}, we divide each document into fixed-length chunks of up to 128 words.
In practice, for \shorttitle and baselines, we identify the retrieved chunks, map them back to their original documents, and return the top-$k$ documents.
We use Normalised Cumulative Discount Gain (nDCG@k) as the evaluation metrics for document retrieval.
Unlike Recall@k, which only indicates the presence of golden documents in the retrieved list, nDCG@k also accounts for both the ranking of retrievals and the relevance judgment of golden documents \cite{thakur2021beir}.
The baselines will be the metrics containing the homogeneous granularity introduced in the previous section, i.e., $s_{q\text{-}d}$, $s_{q\text{-}p}$ and $s_{s\text{-}p}$.

\subsection{Downstream QA Evaluation}
\label{sec:downstream_qa}
As previously mentioned, scientific documents are vital for LLMs due to the rapid advancements in science and the limited availability of such content in training datasets.
To better understand how \shorttitle enhances downstream QA tasks, we implement the \ti{retrieval-then-read} approach on two datasets SciQ and SciFact.
We retrieve and rank the top-$k$ documents based on scores, $s_{q\text{-}d}$ and \shorttitle, then concatenate them to form the context.
During our evaluations, we limit the number of document chunks retrieved to 1 and 3—thus, only the top $k$ documents are injected into the reader model.
We assess the performance by measuring the Exact Match (EM) rate—the proportion of responses where the predicted answer perfectly aligns with the reference answer \cite{kamalloo-etal-2023-evaluating}, denoted as EM@$k$.
Specifically, we utilize LLama-3-8B-Instruct \footnote{\url{https://huggingface.co/meta-llama/Meta-Llama-3-8B-Instruct}} \cite{DBLP:journals/corr/abs-2307-09288} as the reader model.
We take the original query-doc retrieval setup, i.e., retrieval based on $s_{q\text{-}d}$, as the baseline.
Please refer to Appendix \ref{appendix:downstream_task} for more details.

\section{Results} \label{sec:results}
This section analyzes the impact of mixed-granularity retrieval on document retrieval and downstream applications.
We highlight the effectiveness of our proposed fine-grained and mixed-granularity approaches in enhancing performance across various metrics.

\subsection{Document Retrieval} \label{sec:document_retrieval}
Table \ref{tab:ir} reports the results of document retrieval.
We observe that retrieval by \shorttitle outperforms all single-granularity retrieval with both unsupervised and supervised dense retrievers in most cases.

With unsupervised retrievers, \shorttitle significantly outperforms the query-doc similarity, $s_{q\text{-}d}$, across all five datasets.
There is an average nDCG@5 improvement of $+10.2$ and $+5.1$ ($39.2\%$ and $10.1\%$ relatively) for SimCSE and Contriever, respectively.

With supervised retrievers, improvements associated with \shorttitle are also observed, although they are not as significant as with unsupervised retrievers.
This indicates that \shorttitle effectively narrows the distributional gap between dense retrievers and scientific domains.

\begin{figure}[t]
\centering
\includegraphics[width=\linewidth]{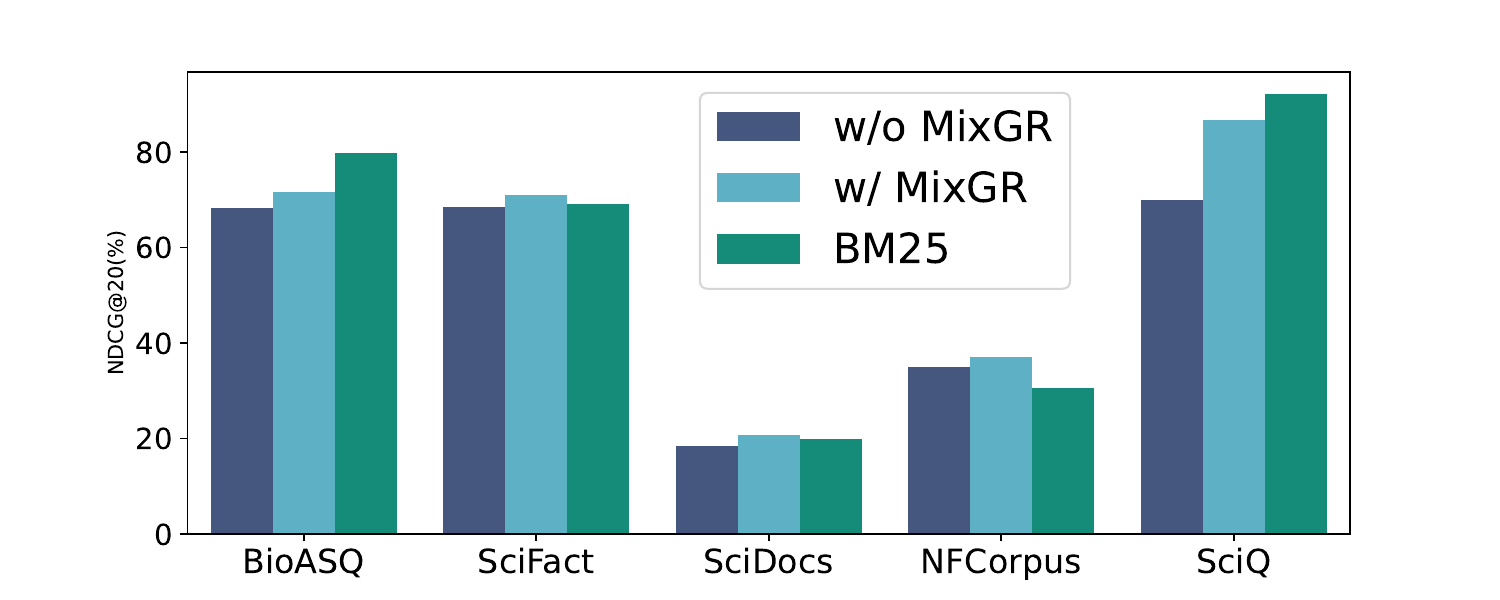}
\caption{Comparison between BM25 and Contriever (w/ and w/o \shorttitle) on nDCG@20: Contriever w/ \shorttitle outperforms BM25 in three out of five datasets.}
\label{fig:bm25}
\end{figure}

\paragraph{Unsupervised retrievers benefit more from \shorttitle than supervised ones.}
Remarkably, with \shorttitle, the unsupervised retriever Contriever outperforms supervised models, as evidenced by its superior average results across five datasets on nCCG@20.
This result is particularly significant given that Contriever typically underperforms compared to TAS-B and GTR when evaluated using traditional query-document similarity measures.
Additionally, the study \cite{thakur2021beir} reveals that sparse retrievers like BM25 often excel over dense retrievers in domain-specific retrieval tasks.
As shown in Figure \ref{fig:bm25}, Contriever outperforms BM25 in three out of five datasets when applied with \shorttitle.
These findings emphasize the substantial enhancements that \shorttitle contributes to unsupervised retrievers within scientific domains.

\paragraph{Finer granularity helps retrieval more.}
Among three metrics within \shorttitle, the query-proposition measurement $s_{q\text{-}p}$ and the subquery-proposition measurement $s_{s\text{-}p}$ show a general advantage over the original query-doc similarity, as highlighted by the \underline{underlined} results in Table \ref{tab:ir}.
The original query-doc metric, $s_{q\text{-}d}$, outperforms the subquery-proposition measurement only when using the retriever TAS-B.
These findings corroborate and expand upon \citet{chen2023dense}, suggesting that finer query-doc similarity measurement significantly improves document retrieval performance.

\subsection{Downstream QA Tasks}
Table \ref{tab:qa} reports the results of scientific question answering when the documents retrieved by \shorttitle are fed into LLMs, i.e. the readers.
It is observed that EM scores achieved with \shorttitle generally surpass those of the baseline across two datasets, six dense retrievers, and multiple numbers of input documents.
This underscores the effectiveness of \shorttitle in enhancing the performance of downstream QA tasks.

\begin{table}[t]
\centering
\resizebox{0.45\textwidth}{!}{%
\begin{tabular}{lc|cc|cc}
\toprule
\multirow{2}{*}{} & \multicolumn{1}{c|}{\multirow{2}{*}{\tf{Setup}}} & \multicolumn{2}{c|}{\tf{SciFact}} & \multicolumn{2}{c}{\tf{SciQ}}  \\
 & & EM@1  & EM@3 & EM@1  & EM@3  \\ 
\midrule
\multicolumn{6}{c}{Unsupervised Dense Retrievers} \\
\midrule
\multirow{2}{*}{\tf{SimCSE}}    & $s_{q\text{-}d}$ & \tf{50.0} & 61.6 & 54.7 & 58.2 \\
& \shorttitle      &  48.3 & \tf{62.8} & \tf{61.3} & \tf{66.4} \\

\midrule
\multirow{2}{*}{\tf{Contriever}} & $s_{q\text{-}d}$ & 63.4 & \tf{75.6} & 53.9 & 63.3 \\
                           & \shorttitle & \tf{64.0} & 70.9 & \tf{61.7} & \tf{66.0} \\

\midrule
\multicolumn{6}{c}{Supervised Dense Retrievers} \\
\midrule
\multirow{2}{*}{\tf{DPR}}       & $s_{q\text{-}d}$ & 51.2 & 59.9 & 52.0 & 57.4  \\
                           & \shorttitle &  \tf{51.7} & \tf{65.7} & \tf{57.4} & \tf{62.5} \\
\midrule
\multirow{2}{*}{\tf{ANCE}}      & $s_{q\text{-}d}$ & 51.7 & 65.1 & 52.7 & 59.4 \\
                           & \shorttitle & \tf{57.6} & \tf{69.2} & \tf{54.7} & \tf{62.9} \\
\midrule
\multirow{2}{*}{\tf{TAS-B}}     & $s_{q\text{-}d}$ & \tf{62.8} & \tf{74.4} & 60.5 & 66.4 \\
                           & \shorttitle & 62.2 & 70.3 & \tf{64.5} & \tf{67.6}  \\
\midrule
\multirow{2}{*}{\tf{GTR}}       & $s_{q\text{-}d}$ & 61.0 & 72.1 & 59.8 & 64.8\\
                           & \shorttitle & \tf{62.8} & \tf{73.8} & \tf{64.1} & \tf{66.0} \\

\bottomrule
\end{tabular}%
}
\vspace{-0.07in}
\caption{
Scientific Question Answering on SciFact and SciQ using Llama-3-8B-Instruct \cite{DBLP:journals/corr/abs-2307-09288}:
the top-1 and 3 document chunks retrieved by retrievers, following the metrics $s_{q\text{-}d}$ and \shorttitle, were fed into the reader. \tf{Bold} indicates the better performance.
}
\label{tab:qa}
\vspace{-1em}
\end{table}

\section{Analysis} \label{sec:analysis}
In this section, we explore the complementary advantages of various similarity metrics across multiple granularities within \shorttitle through an ablation study.
Although the finer-granularity metric, $s_{s\text{-}p}$,  generally enhances performance as previously discussed, it can occasionally result in degradation when compared to original query-document similarity $s_{q\text{-}d}$.
We identify specific conditions under which the finer-granularity metric offers greater benefits.
Previous works \cite{chen2023dense} primarily explored multiple granularities in \ti{documents}. We conduct a controlled experiment to highlight the significance of incorporating multiple granularities in \ti{queries} in the \shorttitle framework, which also validate the generalization of \shorttitle on the retrieval units finer than documents.

\subsection{Ablation Study} \label{sec:ablation_study}

\begin{figure}[t]
\centering
\includegraphics[width=\linewidth]{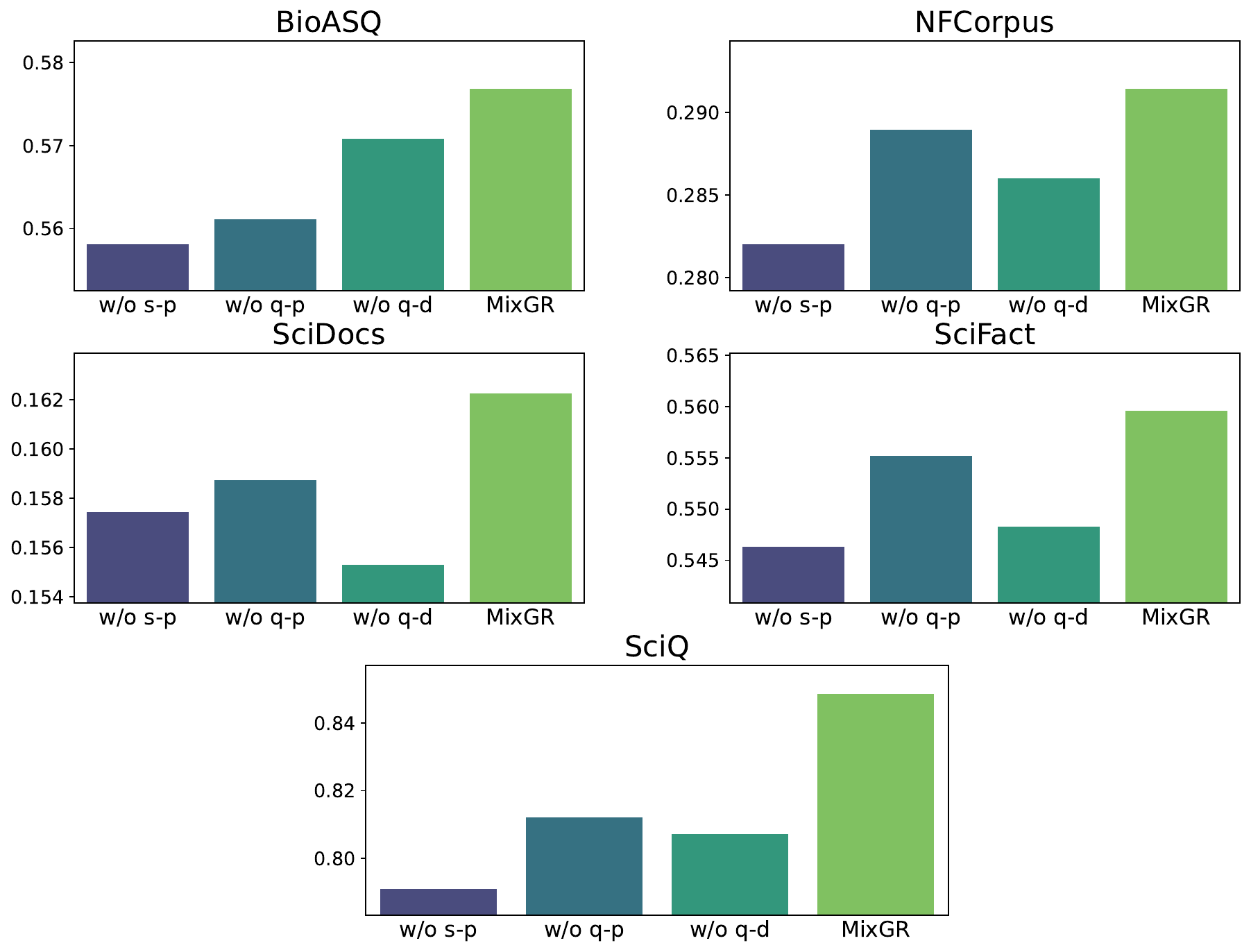}
\caption{Ablation study of \shorttitle on the nDCG@20 metrics averaged on six retrievers: \shorttitle achieves optimal performance when combining these three metrics, indicating their complementary nature.}
\label{fig:ablation}
\end{figure}

In our ablation study, we conducted a systematic evaluation of the impact of various granularity measures—$s_{q\text{-}d}$ (query-doc similarity), $s_{q\text{-}p}$ (query-prop similarity), and $s_{s\text{-}p}$ (subquery-prop similarity)—on the performance of six retrievers.
By individually omitting each of these measures from the calculation of \shorttitle as defined in Equation \ref{eq:rrf}, we assessed the significance of each granularity level.
Specifically, the extent of performance degradation upon removal of a measure indicates its importance; greater degradation suggests higher importance of that particular granularity metric.

As illustrated in Figure \ref{fig:ablation}, the nDCG@20 performance declined across all three setups and datasets, demonstrating that the metrics are complementary to each other.
The degree of performance degradation varied across different configurations, highlighting the importance of each granularity measure. 
Notably, the most significant declines in performance consistently occurred in configurations excluding $s_{q\text{-}d}$ and $s_{s\text{-}p}$.
This observation suggests that $s_{q\text{-}p}$, while beneficial, is probably the \ti{least} critical measure for retrieval tasks in scientific domains.
Please refer to Table \ref{tab:ablation} in Appendix \ref{appendix:ablation} for detailed results.

\subsection{When is finer granularity beneficial?} \label{sec:finer_granularity}

Therefore, to more effectively compare the impacts of \( s_{q\text{-}d} \) and \( s_{s\text{-}p} \), we categorized the \ti{correctly} retrieved pairs (complex query, \footnote{We refer \ti{complex query} as the query containing no fewer than three subqueries.} doc) by \shorttitle in SciFact, using SimCSE, into two distinct groups:

\begin{itemize}
    \item \( r_{q\text{-}d} \succ r_{s\text{-}p} \): The query-doc rank of $s_{q\text{-}d}$ is higher than the subquery-prop rank of $s_{s\text{-}p}$;
    \item \( r_{q\text{-}d} \prec r_{s\text{-}p} \): The query-doc rank of $s_{q\text{-}d}$ is lower than the subquery-prop rank of $s_{s\text{-}p}$.
\end{itemize}

\begin{figure}[t]
\centering
\includegraphics[width=0.9\linewidth]{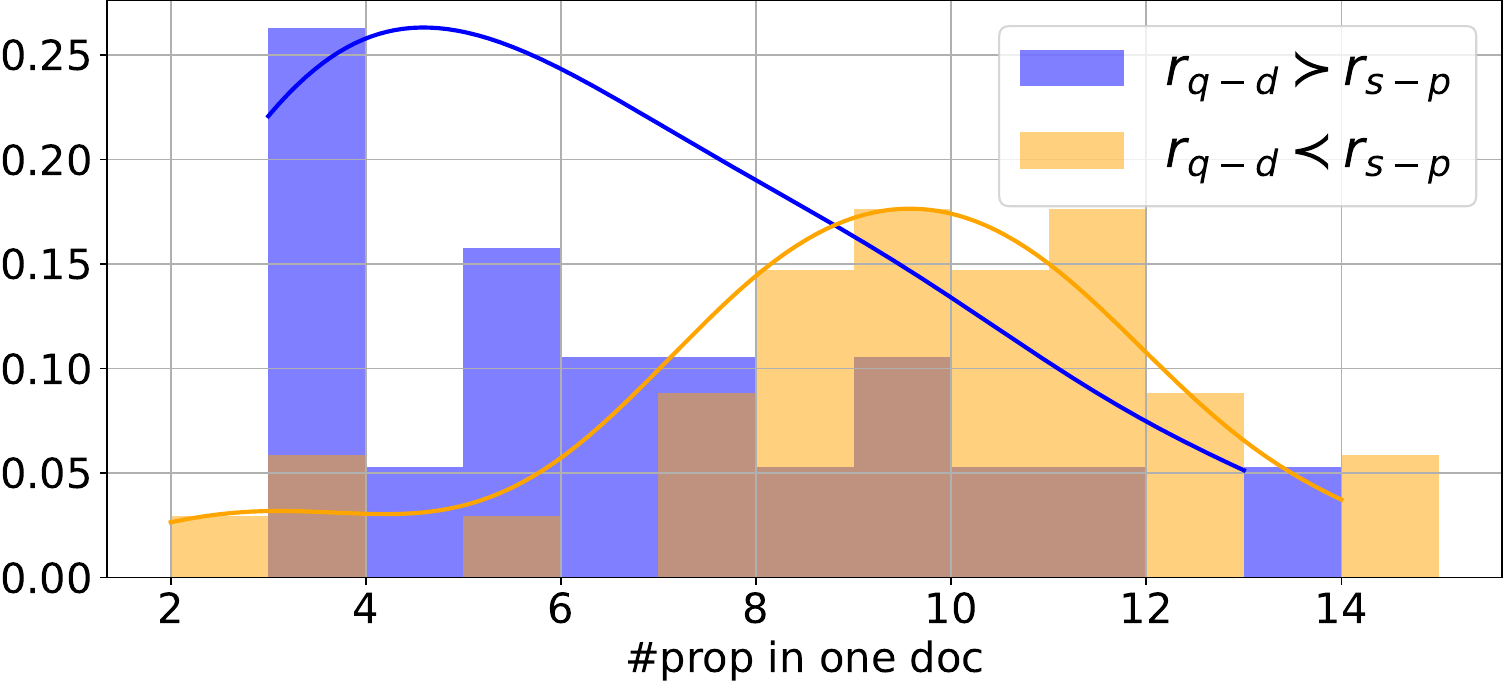}
\caption{Distribution of proposition number within documents in two sets. There are more propositions within document when \( r_{q\text{-}d} \prec r_{s\text{-}p} \) than \( r_{q\text{-}d} \succ r_{s\text{-}p} \). }
\label{fig:proposition_dist}
\end{figure}

Upon analyzing the number of propositions in documents, a significant pattern emerges:
based on the distributions present in Figure \ref{fig:proposition_dist}, the number of propositions in  \( r_{q\text{-}d} \prec r_{s\text{-}p} \) is generally higher than in \( r_{q\text{-}d} \succ r_{s\text{-}p} \).
This underscores the importance of incorporating finer units within documents, especially for those containing more propositions, and suggests potential degradation in dense retrievers when handling such documents.
For other retrievers' results, please refer to Appendix \ref{appendix:two_sets}.

\subsection{\shorttitle on Proposition Retrieval} \label{sec:prop_retrieval}
Previous sections present the effectiveness of \shorttitle on scientific document retrieval.
While previous works \cite{chen2023dense} focus on finer document granularity, we specifically assess \shorttitle on the proposition as the retrieval units.
This controlled study highlights the benefits of \shorttitle, which incorporates different granularities within queries and documents, in general text retrieval beyond document-level granularity.

For a given query $\mathrm{q}$ and a proposition $\mathrm{p}$, the conventional similarity is denoted by $s^p_{q\text{-}p} \equiv s(\mathrm{q}, \mathrm{p})$.
When the query is further broken down into multiple sub-queries, we introduce a finer granularity measure, $s^p_{s\text{-}p}$, which is defined as the maximum similarity between these sub-queries and the proposition.
$s^p_{s\text{-}p}$ is mathematically defined as follows:

\begin{align}
    s^p_{s\text{-}p}(\mathrm{q}, \mathrm{p}) = \max_{i=1,...,M}{ \{s(q_i, \mathrm{p})\} }.
    \label{eq:sub-prop-on-prop}
\end{align}

Therefore, the merged score by RRF, $s^p_f(\mathrm{q}, \mathrm{p})$, is calculated as:
\begin{align}
    s^p_f(\mathrm{q}, \mathrm{p}) = \frac{1}{1 + r^p_{q\text{-}p}(\mathrm{q}, \mathrm{p})} + \frac{1}{1 + r^p_{s\text{-}p}(\mathrm{q}, \mathrm{p})},
    \label{eq:rrf-prop}
\end{align}

where $r^p_{q\text{-}p}$ and $r^p_{s\text{-}p}$ signify the rank of the retrieve results by $s^p_{q\text{-}p}$ and $s^p_{s\text{-}p}$, respectively.

\begin{figure}[t]
\centering
\includegraphics[width=\linewidth]{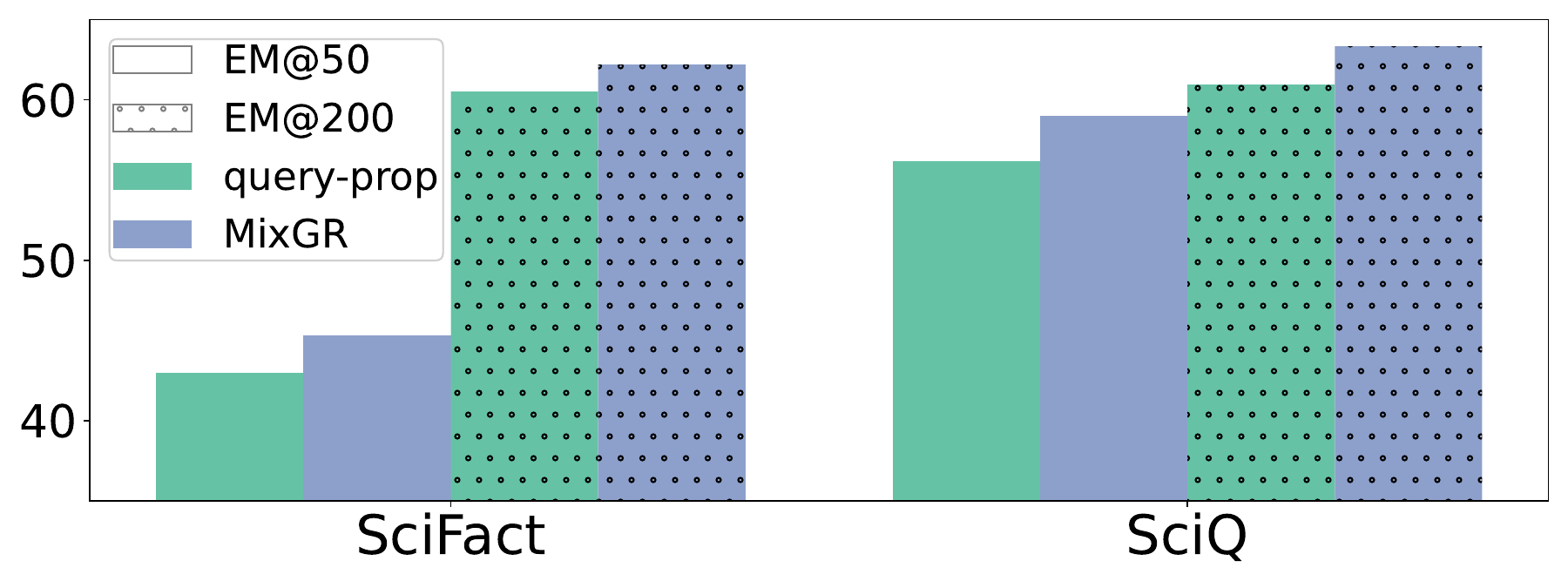}
\caption{Proposition retrieval with \shorttitle:
We evaluate Exact Match of LLama-3-8B-Instruct on SciFact and SciQ with the first 50 and 200 words of propositions, i.e., EM@50 and EM@200, retrieved by SimCSE as the context. Please refer to Table \ref{tab:qa_proposition} for other retrievers in Appendix \ref{appendix:proposition}.}
\label{fig:proposition_extension}
\end{figure}

Following \( s^p_{q\text{-}p}(\mathrm{q}, \mathrm{p}) \) and \( s^p_f(\mathrm{q}, \mathrm{p}) \), we input the first 50 and 200 words in propositions retrieved with SimCSE on SciFact and SciQ into the reader LLama-3-8B-Instruct. This process adheres to the same setups outlined in \S \ref{sec:downstream_qa}.
As shown in Figure \ref{fig:proposition_extension}, the performance advance observed with mixed-granularity retrieval on propositions, compared to the original query-prop similarity, demonstrates the effectiveness of using mixed-granularity in retrieval.
This substantiates the generalizability of \shorttitle beyond document-level granularity.
Please refer to Appendix \ref{appendix:proposition} for details.

\subsection{Prospect: Adaptive \shorttitle}
Here, we outline potential future research directions.
In \S \ref{sec:ablation_study}, we observed the complementary nature of retrieval results achieved using different granularities.
Additionally, as discussed in \S \ref{sec:finer_granularity}, we noted a distinct pattern where retrieval guided by a specific granularity outperforms others.
These findings indicate that metrics based on different granularities each have relatively distinct strengths in specific contexts, presenting opportunities for further exploration.
Unlike the non-parametric method of fusion by RRF, which overlooks the relative importance of components, an adaptive approach could enhance fusion and, consequently, improve retrieval performance with dense retrievers--a prospect we aim to explore in future research.

\section{Conclusion}
In this work, we identify key challenges for dense retrievers in scientific document retrieval, namely domain shift and query-document complexity.
In response, we propose a zero-shot approach, \shorttitle, that utilizes atomic components in queries and documents to calculate their similarity with greater nuance.
We then use Reciprocal Rank Fusion (RRF) to integrate these metrics, modeling query-doc similarity at different granularities into a unified score that enhances document retrieval.

Our experiments demonstrate that \shorttitle significantly enhances the existing dense retriever on document retrieval within the scientific domain.
Moreover, \shorttitle has proven beneficial for downstream applications such as scientific QA.
The analysis reveals a synergistic relationship among the components of \shorttitle, and suggests evolving our non-parametric fusion framework into a more general method as a future research direction.

\section*{Limitations}
Our work explores retrieval guided by an integral metric that incorporates various levels of granularity. We identify several limitations in our approach:
(1) \textit{Coverage of Retrievers}: Our study categorizes dense retrievers into supervised and unsupervised models, yet all utilize a dual-encoder structure.
Future studies could include a more diverse array of retriever architectures.
(2) \textit{Coverage of Domains}:
While our main focus is on the scientific domain, and we extend to three additional domains in Appendix \ref{appendix:other_domains}, there are still many domains we have not explored.
(3) \textit{Languages}: Our research is limited to an English corpus.
The applicability of \shorttitle in multilingual contexts also deserves further validation and exploration.
(4) Based on our error analysis in Appendix \ref{appendix:propositioner}, there is potential for improving the tool used for query and corpus decomposition.

\section*{Ethical Statements}
We foresee no ethical concerns and potential risks in our work.
All of the retrieval models and datasets are open-sourced, as shown in Table \ref{tab:tools} in Appendix \ref{appendix:artifacts}.
The LLMs we applied in the experiments are also publicly available.
Given our context, the outputs of LLMs are unlikely to contain harmful and dangerous information.

\section*{Acknowledgement}
We thank Kexin Wang, Tim Baumgärtner, and Sherry Tongshuang Wu for the discussion on an early draft of this work.

Fengyu Cai is funded by the German Federal Ministry of Education and Research and the Hessian Ministry of Higher Education, Research, Science, and the Arts within their joint support of the National Research Center for Applied Cybersecurity ATHENE.
Xinran Zhao is funded by the ONR Award N000142312840.

\bibliographystyle{acl_natbib}
\bibliography{custom, anthology, anthology_p1}

\clearpage
\onecolumn
\appendix
\part*{Appendix}

\section{Datasets} \label{appendix:datasets}
\subsection{General Setups}
Different from the setup of the original dataset, we split one document into several chunks with a maximum of 128 words. This is because some dense retrievers such as DPR \cite{karpukhin-etal-2020-dense} have the requirement of maximum input.
Too long inputs will be overflow, leading to the loss of information.
The chunk selected can be used to locate the document in the original dataset during the evaluation.
Specifically, for SciQ, we reformulate the dataset from a QA task to a retrieval task.
Originally, this task aims to answer scientific questions given the context.
We collect the contexts in training, validation and test sets as the corpus.

Also, we will explain our motivation of focusing the queries containing subqueries:

\begin{itemize}[leftmargin=*, itemsep=0em]
    \item \citet{chen2023dense} have studied the advantage of using propositions, i.e., the atomic units within documents, as the retrieval units for a complete query. And \shorttitle will not affect the retrieval results of single-subquery queries.
    \item In this work, we highlight the advantages of mixed-granularity retrieval that incorporates finer units in both queries and documents. Queries containing multiple subqueries are particularly well-suited to our research problem, as they will have different combinations with the documents.
\end{itemize}

\subsection{Specific Setups}

Here, we would specify the experimental setups for each dataset:

\begin{itemize}[leftmargin=*, itemsep=0em]
    \item BioASQ: In order to decrease the computational cost, we randomly sample the abstracts from the original corpus covered by BEIR \cite{thakur2021beir}, which contains \numprint{14914603} abstracts.
    \item NFCorpus: The query that we apply here is the description version including training, validation, and test sets, instead of the keyword. These queries are not covered by BEIR \cite{thakur2021beir}. Please refer to the original datasets of NFCorpus \cite{boteva2016full}.
    \item SciFact: We apply the dataset covered by BEIR \cite{thakur2021beir}, using both training and test splits.
    \item SciDocs: We use the dataset covered by BEIR \cite{thakur2021beir}, using the test split.
    \item SciQ: We use the queries in the test sets, and collect the contexts in training, validation, and test sets as the corpus.
\end{itemize}

For the documents (or propositions) of all the datasets, the format of the text is the concatenation of the title and the content. 

\begin{table}[h]
    \centering
    \resizebox{\textwidth}{!}{%
    \begin{tabular}{lccccc}
        \toprule
        \textbf{Statistic} & \tf{BioASQ} \cite{tsatsaronis2015overview} & \tf{NFCorpus} \cite{boteva2016full} & \tf{SciDocs} \cite{cohan-etal-2020-specter} & \tf{SciFact} \cite{wadden-etal-2020-fact} & \tf{SciQ} \cite{welbl-etal-2017-crowdsourcing} \\
        \midrule
        \#Query & \numprint{3743} & \numprint{1016} & \numprint{1000} & \numprint{1109} & \numprint{884} \\
        \#Multi-subquery queries & \numprint{387} & \numprint{641} & \numprint{205} & \numprint{283} & \numprint{252} \\
        \#Subqueries & \numprint{821} & \numprint{3337} & \numprint{522} & \numprint{614} & \numprint{874} \\ 
        \midrule
        \#Documents & \numprint{225362} & \numprint{3633} & \numprint{25657} & \numprint{5183} & \numprint{12241} \\
        \#Propositions & \numprint{3551816} & \numprint{67110} & \numprint{351802} & \numprint{87190} & \numprint{91635} \\
        \bottomrule
    \end{tabular}
    }
    \caption{Statistics for the BioASQ, NFCorpus, SciDocs, SciFact, and SciQ datasets. Please note that these statistics have been adjusted to exclude any obvious decomposition errors.}
    \label{tab:dataset_stats}
\end{table}

\section{Query and Document Decomposition} \label{appendix:examples}

Here, we will complement the necessary information regarding the query and document decomposition.

\subsection{Subquery and Proposition Examples}
Here, we present examples of subqueries and propositions decomposed from the documents.
The example is the decomposition of the example in Figure \ref{fig:both_figures}.

\begin{tcolorbox}
Query: Citrullinated proteins externalized in neutrophil extracellular traps act indirectly to perpetuate the inflammatory cycle via induction of autoantibodies.
\begin{itemize}
    \item Subquery-0: Citrullinated proteins are externalized in neutrophil extracellular traps.
    \item Subquery-1: Citrullinated proteins act indirectly to perpetuate the inflammatory cycle.
    \item Subquery-2: The inflammatory cycle is perpetuated via induction of autoantibodies.
\end{itemize}

\vspace{1em}

Document: RA sera and immunoglobulin fractions from RA patients with high levels of ACPA and/or rheumatoid factor significantly enhanced NETosis, and the NETs induced by these autoantibodies displayed distinct protein content. Indeed, during NETosis, neutrophils externalized the citrullinated autoantigens implicated in RA pathogenesis, and anti-citrullinated vimentin antibodies potently induced NET formation. Moreover, the inflammatory cytokines interleukin-17A (IL-17A) and tumor necrosis factor-$\alpha$ (TNF-$\alpha$) induced NETosis in RA neutrophils. In turn, NETs significantly augmented inflammatory responses in RA and OA synovial fibroblasts, including induction of IL-6, IL-8, chemokines, and adhesion molecules. These observations implicate accelerated NETosis in RA pathogenesis, through externalization of citrullinated autoantigens and immunostimulatory molecules that may promote aberrant adaptive and innate immune responses in the joint and in the periphery, and perpetuate pathogenic mechanisms in this disease.

\begin{itemize}
    \item Proposition-0: RA sera and immunoglobulin fractions from RA patients with high levels of ACPA and/or rheumatoid factor significantly enhanced NETosis.
    \item Proposition-1: NETs induced by these autoantibodies displayed distinct protein content.
    \item Proposition-2: During NETosis, neutrophils externalized the citrullinated autoantigens implicated in RA pathogenesis.
    \item Proposition-3: Anti-citrullinated vimentin antibodies potently induced NET formation.
    \item Proposition-4: Interleukin-17A (IL-17A) and tumor necrosis factor- (TNF-) induced NETosis in RA neutrophils.
    \item Proposition-5: NETs significantly augmented inflammatory responses in RA and OA synovial fibroblasts.
    \item Proposition-6: NETs inducing IL-6, IL-8, chemokines, and adhesion molecules occurred in RA and OA synovial fibroblasts.
    \item Proposition-7: These observations implicate accelerated NETosis in RA pathogenesis.
    \item Proposition-8: NETosis externalizes citrullinated autoantigens and immunostimulatory molecules.
    \item Proposition-9: NETosis may promote aberrant adaptive and innate immune responses in the joint and in the periphery.
    \item Proposition-10: NETosis may perpetuate pathogenic mechanisms in RA.
\end{itemize}

\end{tcolorbox}

\subsection{Remarks on \ti{Propositioner}} \label{appendix:propositioner}
During our manual check on the decomposition results of \ti{propositioner} \cite{chen2023dense}, we find the following potential flaws.

(1) Wrong logic during decomposition:
\begin{tcolorbox}
\ti{Query}: Identification of Design Elements for a Maturity Model for Interorganizational Integration: A Comparative Analysis

$\rightarrow$
\ti{Subqueries}: ['Identification of Design Elements for a Maturity Model for Interorganizational Integration.', 'A Comparative Analysis is used for identifying design elements.']
\end{tcolorbox}

(2) Hallucination:

\begin{tcolorbox}
\ti{Query}: Bigger ocean waves and waves that carry more sediment cause a greater extent of what?

$\rightarrow$
\ti{Subqueries}: ['Bigger ocean waves cause a greater extent of erosion.', 'Waves that carry more sediment cause a greater extent of erosion.']
\end{tcolorbox}

(3) Information loss:

\begin{tcolorbox}
\ti{Query}: The reduction was 1.6 ± 1.6 in controls. ...

$\rightarrow$
\ti{Subqueries}: ['The reduction in migraine headache was 1.6  1.6 in controls.', ...]
\end{tcolorbox}

We find that the proposition will convert the questions to declarative sentences during decomposition.
This may stem from the fact that its training corpus is Wikipedia, where a small portion of sentences are questions.
Still, we find that \ti{propositioner} can still decompose question-style queries, as shown in the following example:

\begin{tcolorbox}
\ti{Query}: What is the purpose of bright colors on a flower's petals?

$\rightarrow$
\ti{Subqueries}: ["The purpose of bright colors on a flower's petals is unknown."]
\end{tcolorbox}

What is more, the propositioner may decompose the query to a sequence of single characters, but it is very rare: there are only four cases out of 4009 queries, i.e., around 0.1 \% rate for this type of error.

\subsection{Human Evaluation on Query and Document Decomposition}
As mentioned in \S \ref{sec:query-document}, we evaluate the decomposition outputs by \ti{propositioner} \cite{chen2023dense}, 100 samples for both query and document decomposition.
Concretely, we ask three students at the post-graduate levels to evaluate the results, who are paid above the local minimum hourly wage.
The instruction is shown below:

\begin{tcolorbox}
Propositions in documents (or subqueries in queries) are defined as follows:
\begin{itemize}[leftmargin=*, itemsep=0em]
    \item Each proposition conveys a distinct semantic unit, collectively expressing the complete meaning.
    \item Propositions should be atomic and indivisible.
    \item According to \citet{choi2021decontextualization}, propositions should be contextualized and self-contained, including all necessary text information such as coreferences for clear interpretation.
\end{itemize}

Given the document (query) and the corresponding propositions (subqueries) generated by the model, please check whether the document (query) has been correctly decomposed.

Please write \ti{1} as correct, and \ti{0} as incorrect.
\end{tcolorbox}

\section{Retrievers Models} \label{appendix:retrievers}
Table \ref{tab:model_checkpoints} presents the dense retrievers applied in the experimental section, i.e., \S \ref{sec:experiments}.

\begin{table}[t]
    \centering
    \resizebox{\textwidth}{!}{%
    \begin{tabular}{l >{\ttfamily}l}
        \toprule
        \textbf{Model} & \textbf{HuggingFace Checkpoint} \\
        \midrule
        SimCSE \cite{gao-etal-2021-simcse} & princeton-nlp/unsup-simcse-bert-base-uncased \\
        Contriever \cite{izacard2022unsupervised} & facebook/contriever \\
        DPR \cite{karpukhin-etal-2020-dense} & facebook/dpr-ctx\_encoder-multiset-base \\
        & facebook/dpr-question\_encoder-multiset-base \\
        ANCE \cite{xiong2021approximate} & castorini/ance-dpr-context-multi \\
        & castorini/ance-dpr-question-multi \\
        TAS-B \cite{hofstatter2021efficiently} & sentence-transformers/msmarco-distilbert-base-tas-b \\
        GTR \cite{ni-etal-2022-large} & sentence-transformers/gtr-t5-base \\
        \bottomrule
    \end{tabular}
    }
    \caption{Model checkpoints released on HuggingFace. For DPR and ANCE, two different models encode the context and query.}
    \label{tab:model_checkpoints}
\end{table}

\newblock

\section{Offline Indexing} \label{appendix:indexing}

The \texttt{pyserini} and \texttt{faiss} libraries were employed to convert retrieval units into embeddings.
We leveraged GPUs for encoding these text units in batches with a batch size of 64 and a floating precision of f16.
Following the preprocessing of these embeddings, all experiments conducted involved the utilization of an exact search method for inner products using \texttt{faiss.IndexFlatIP},

\section{Downstream Tasks} \label{appendix:downstream_task}
The templates of LLama for downstream QA tasks, i.e., SciFact and SciQ, are listed as follows.
For SciQ, we convert it from multiple choice question answering to open question answering.

\begin{tcolorbox}
    Given the knowledge source: \ti{context} \textbackslash\textbackslash n Question: \ti{query} \textbackslash\textbackslash n Reply with one phrase. \textbackslash\textbackslash n Answer:

\end{tcolorbox}

As SciFact is a fact-checking task, we here check whether LLMs can predict the relationship between the context and the claim.
The template of SciFact is shown as follows:
\begin{tcolorbox}
Context: \{\ti{context}\} \textbackslash\textbackslash n Claim: \{\ti{query}\} \textbackslash\textbackslash n For the claim, the context is supportive, contradictory, or not related? \textbackslash\textbackslash n Options: (A) Supportive (B) Contradictory (C) Not related \textbackslash\textbackslash n Answer:")
\end{tcolorbox}

\section{Detailed Results}
\subsection{Ablation Study} \label{appendix:ablation}
As discussed in \S \ref{sec:ablation_study}, we remove the component, i.e., query-doc similarity, query-prop similarity, or subquery-prop similarity, and assess the corresponding performance compared with \shorttitle.
In Table \ref{tab:ablation}, it is observed that \shorttitle outperforms all its components.

\begin{table*}[h]
\centering
\resizebox{0.95\textwidth}{!}{%
\begin{tabular}{lc|cc|cc|cc|cc|cc|cc}
\toprule
\multirow{2}{*}{\tf{Retriever}} & \multicolumn{1}{c|}{\multirow{2}{*}{\tf{Setup}}} & \multicolumn{2}{c|}{\tf{BioASQ}}  &  \multicolumn{2}{c|}{\tf{NFCorpus}} & \multicolumn{2}{c|}{\tf{SciDocs}} & \multicolumn{2}{c|}{\tf{SciFact}} & \multicolumn{2}{c|}{\tf{SciQ}} & \multicolumn{2}{c}{\tf{Avg.}} \\
 & & ND@5  & ND@20 & ND@5  & ND@20 & ND@5  & ND@20 & ND@5  & ND@20 & ND@5  & ND@20 & ND@5  & ND@20 \\ 
\midrule
\multicolumn{14}{c}{Unsupervised Dense Retrievers} \\
\midrule
\multirow{4}{*}{\tf{SimCSE}}    & w/o $s_{s\text{-}p}$ & 25.0 & 26.1 &19.6 & 16.0& 8.7 & 11.5& 32.3 & 37.0& 76.1 & 78.0& 32.3 & 33.7 \\
& w/o $s_{q\text{-}p}$ &27.6 & 28.6& 21.4 & 17.4& 8.5 & 11.6& 33.1 & 37.4& 77.9 & 79.6& 33.7 & 34.9 \\
& w/o $s_{q\text{-}d}$ &32.4 & 32.4& 22.8 & 18.6& 8.5 & 11.9& 33.9 & 39.0& 80.7 & 82.2& 35.7 & 36.8 \\
& \shorttitle      &30.7 & 31.3& 22.3 & 18.1& 9.1 & 12.2& 34.8 & 39.8& 84.0 & 85.5& 36.2 & 37.4 \\

\midrule
\multirow{4}{*}{\tf{Contriever}} &  w/o $s_{s\text{-}p}$ &66.5 & 70.5 & 43.6 & 36.2& 14.8 & 20.0& 65.6 & 69.9& 78.0 & 80.1& 53.7 & 55.3 \\
                           & w/o $s_{q\text{-}p}$ & 66.9 & 71.2& 43.0 & 36.6& 14.6 & 20.1& 66.3 & 70.8& 81.6 & 83.3&54.5 & 56.4 \\
                           & w/o $s_{q\text{-}d}$ &66.1 & 70.4 & 43.2 & 36.3& 14.7 & 20.0& 65.0 & 69.5& 83.3 & 84.8& 54.5 & 56.2 \\
                           & \shorttitle & 67.0 & 71.7& 44.0 & 37.1& 15.5 & 20.7& 66.4 & 71.0& 85.2 & 86.7& 55.6 & 57.5 \\
\midrule
\multicolumn{14}{c}{Supervised Dense Retrievers} \\
\midrule
\multirow{4}{*}{\tf{DPR}}       &  w/o $s_{s\text{-}p}$ &43.0 & 44.3& 26.5 & 21.9& 8.2 & 11.2& 35.0 & 40.8& 66.6 & 69.9& 35.9 & 37.6 \\
                           & w/o $s_{q\text{-}p}$ & 42.9 & 44.5& 27.5 & 22.8& 7.5 & 11.2& 38.3 & 42.4& 71.0 & 73.1& 37.5 & 38.8 \\
                           & w/o $s_{q\text{-}d}$ & 44.0 & 45.0 & 26.6 & 22.2& 8.0 & 11.2& 38.0 & 42.1& 69.5 & 72.2& 37.2 & 38.5 \\
                           & \shorttitle & 44.6 & 45.8& 27.7 & 22.9& 8.2 & 11.5& 39.4 & 43.6& 73.6 & 76.1& 38.7 & 40.0 \\
\midrule
\multirow{4}{*}{\tf{ANCE}}      &  w/o $s_{s\text{-}p}$ & 52.7 & 52.9 & 30.7 & 25.2& 10.0 & 13.7& 45.8 & 48.9& 69.0 & 72.0& 41.7 & 42.6 \\
                           & w/o $s_{q\text{-}p}$ & 51.4 & 52.8&  32.0 & 26.2& 9.0 & 13.4& 46.8 & 50.4& 71.3 & 73.9& 42.1 & 43.3 \\
                           & w/o $s_{q\text{-}d}$ &  52.9 & 54.1 & 30.8 & 25.1& 8.8 & 13.4& 44.9 & 48.6& 67.8 & 70.1& 41.0 & 42.3 \\
                           & \shorttitle & 53.4 & 54.7& 31.9 & 25.9& 9.6 & 14.1& 46.8 & 49.9& 74.4 & 76.8& 43.2 & 44.3 \\
\midrule
\multirow{4}{*}{\tf{TAS-B}}     &  w/o $s_{s\text{-}p}$ & 71.3 & 73.2&42.9 & 34.7& 13.8 & 19.2& 61.4 & 66.7& 86.7 & 87.0& 55.2 & 56.2 \\
                           & w/o $s_{q\text{-}p}$ & 70.1 & 72.6&42.9 & 34.9& 13.8 & 19.6& 63.2 & 67.3& 88.3 & 88.8& 55.7 & 56.7 \\
                           & w/o $s_{q\text{-}d}$ & 70.0 & 72.6& 42.7 & 34.5& 13.6 & 18.8& 62.1 & 65.3& 85.2 & 85.9& 54.8 & 55.4 \\
                           & \shorttitle & 71.7 & 74.0& 43.6 & 35.2& 14.0 & 19.6& 62.7 & 66.9& 90.5 & 91.0& 56.5 & 57.3 \\
\midrule
\multirow{4}{*}{\tf{GTR}}       &  w/o $s_{s\text{-}p}$ & 66.5 & 67.8& 43.2 & 35.2& 13.4 & 18.9& 60.9 & 64.5& 87.2 & 87.5& 54.2 & 54.8 \\
                           & w/o $s_{q\text{-}p}$ & 65.4 & 67.0& 43.0 & 35.5& 13.8 & 19.5& 60.6 & 64.7& 88.4 & 88.5& 54.2 & 55.0 \\
                           & w/o $s_{q\text{-}d}$ & 66.2 & 68.0& 42.4 & 34.9& 12.6 & 18.0& 61.5 & 64.4& 89.0 & 89.3& 54.3 & 54.9 \\
                           & \shorttitle &66.8 & 68.6& 43.3 & 35.6& 13.6 & 19.2& 60.9 & 64.5& 92.9 & 93.0& 55.5 & 56.2 \\
\bottomrule
\end{tabular}%
}
\vspace{-0.07in}
\caption{
Ablation study (nDCG@$k$ = 5, 20 in percentage, abbreviated as ND@$k$): We evaluated five distinct scientific retrieval datasets using two unsupervised and four supervised retrievers.
The retrieval results were compared using various metrics: \shorttitle w/o $s_{s\text{-}q}$, \shorttitle w/o $s_{q\text{-}p}$, \shorttitle w/o $s_{s\text{-}p}$, and \shorttitle, as detailed in \S \ref{sec:method:multi-granularity}.
}
\label{tab:ablation}
\end{table*}

\subsection{\shorttitle for Propositional Retrieval} \label{appendix:proposition}
Here, we evaluate \shorttitle on the retrieval units beyond documents, e.g., propositions, which Table \ref{tab:qa_proposition} present.
We observe that \shorttitle can outperform the previous document retrieval based on the similarity between query and proposition, on proposition retrieval, as discussed in \S \ref{sec:prop_retrieval}.

\begin{table}[h]
\centering
\resizebox{0.5\textwidth}{!}{%
\begin{tabular}{lc|cc|cc}
\toprule
\multirow{2}{*}{} & \multicolumn{1}{c|}{\multirow{2}{*}{\tf{Setup}}} & \multicolumn{2}{c|}{\tf{SciFact}} & \multicolumn{2}{c}{\tf{SciQ}}  \\
 & & EM@50  & EM@200 & EM@50  & EM@200  \\ 
\midrule
\multicolumn{6}{c}{Unsupervised Dense Retrievers} \\
\midrule
\multirow{2}{*}{SimCSE}    & \( s_{q\text{-}d} \) & 43.0 & 60.5 & 56.2 & 60.9 \\
                           & \shorttitle      & \tf{45.3} & \tf{62.2} & \tf{59.0} & \tf{63.3} \\

\midrule
\multirow{2}{*}{Contriever} & \( s_{q\text{-}d} \) & \tf{49.4} & 67.4 & 56.2 & \tf{62.9} \\
                           & \shorttitle & 47.7 & \tf{71.5} & \tf{57.4} & 62.5 \\

\midrule
\multicolumn{6}{c}{Supervised Dense Retrievers} \\
\midrule
\multirow{2}{*}{DPR}       & \( s_{q\text{-}d} \) & 49.4 & 56.4 & 55.5 & 60.2 \\
                           & \shorttitle & \tf{52.3} & \tf{59.9} & \tf{59.0} & \tf{60.9} \\
\midrule
\multirow{2}{*}{ANCE}      & \( s_{q\text{-}d} \) & \tf{47.1} & 61.6 & 53.9 & \tf{60.5} \\
                           & \shorttitle & 45.9 & \tf{66.9} & \tf{55.5} & 59.8 \\
\midrule
\multirow{2}{*}{TAS-B}     & \( s_{q\text{-}d} \) & 50.0 & \tf{69.8} & 56.2 & 60.9 \\
                           & \shorttitle & \tf{52.3} & 68.0 & \tf{58.2} & \tf{62.9} \\
\midrule
\multirow{2}{*}{GTR}       & \( s_{q\text{-}d} \) & 41.9 & \tf{66.3} & 60.2 & 63.7 \\
                           & \shorttitle & \tf{45.9} & 63.4 & \tf{60.9} & \tf{65.2} \\
\bottomrule
\end{tabular}%
}
\caption{
Scientific Question Answering (Exact Match) was conducted using LLama-3 \cite{DBLP:journals/corr/abs-2307-09288} with propositions retrieved by six retrievers. Here, EM@50 and EM@200 have been reported, where the first 50 and 200 words are fed into the reader models. \tf{Bold} indicates superior performance, and it is observed that retrieval using \shorttitle on proposition units generally outperforms the baseline.
}
\label{tab:qa_proposition}
\end{table}

\subsection{Advantageous pattern for finer granularity measurement} \label{appendix:two_sets}
In Table \ref{tab:two_sets}, we can notice the average number of propositions in $r_{q\text{-}d} \prec r_{s\text{-}p}$ is more than $r_{q\text{-}d} \succ r_{s\text{-}p}$. This shows that the finer granularity can better deal with the documents with more propositions than the original query-document similarity.

\subsection{Document retrieval for queries containing only a single subquery} \label{appendix:single_subquery}
In this experiment, we demonstrate the impact of \shorttitle on queries that contain only a single subquery.
Unlike queries with multiple subqueries, \shorttitle omits the similarity measurement from the perspectives of subqueries and propositions, $s_{s\text{-}p}$, during RRF.
The result, presented in Table \ref{tab:ir_single}, includes the result of document retrieval for the queries including either one single or multiple subqueries and illustrates the effectiveness of \shorttitle across a general query format.

\begin{table*}[h]
\centering
\resizebox{0.95\textwidth}{!}{%
\begin{tabular}{lc|cc|cc|cc|cc|cc|cc}
\toprule
\multirow{2}{*}{\tf{Retriever}} & \multicolumn{1}{c|}{\multirow{2}{*}{\tf{Setup}}} &  \multicolumn{2}{c|}{\tf{BioASQ}} & \multicolumn{2}{c|}{\tf{NFCorpus}} & \multicolumn{2}{c|}{\tf{SciDocs}} & \multicolumn{2}{c|}{\tf{SciFact}} & \multicolumn{2}{c|}{\tf{SciQ}} & \multicolumn{2}{c}{\tf{Avg.}} \\
 & & ND@5  & ND@20 &ND@5  & ND@20 & ND@5  & ND@20 & ND@5  & ND@20 & ND@5  & ND@20 & ND@5 & ND@20  \\ 
\midrule
\multicolumn{14}{c}{Unsupervised Dense Retrievers} \\
\midrule
\multirow{2}{*}{\tf{SimCSE}}
& $s_{q\text{-}d}$ & 17.7 & 17.9& 14.5 & 12.1& 27.9 & 31.9& 4.9 & 7.1& 50.8 & 56.5& 23.2 & 25.1 \\
& \shorttitle      & \tf{26.2} & \tf{27.1}& \tf{19.8} & \tf{16.2}& \tf{34.8} & \tf{39.1}& \tf{6.7} & \tf{9.5}& \tf{70.1} & \tf{73.0} & \tf{31.5} & \tf{33.0} \\
\midrule
\multirow{2}{*}{\tf{Contriever}}
& $s_{q\text{-}d}$ & 62.0 & 64.4& 39.9 & 33.1& 66.0 & 69.1& 12.4 & 17.2& 56.4 & 60.5& 47.3 & 48.9 \\
& \shorttitle & \tf{64.5} & \tf{67.7} & \tf{41.4} & \tf{34.7} & \tf{67.0} & \tf{70.7} & \tf{13.2} & \tf{18.5} & \tf{75.3} & \tf{77.7} & \tf{52.3} & \tf{53.9} \\
\midrule
\multicolumn{14}{c}{Supervised Dense Retrievers} \\
\midrule
\multirow{2}{*}{\tf{DPR}}
& $s_{q\text{-}d}$ & 36.0 & 35.9& 23.8 & 19.7& 35.9 & 40.5& 6.0 & 8.8& 51.2 & 56.3& 30.6 & 32.2 \\
& \shorttitle & \tf{41.0} & \tf{41.6} & \tf{26.1} & \tf{21.5} & \tf{40.0} & \tf{44.1} & \tf{6.9} & \tf{9.9} & \tf{63.2} & \tf{67.0} & \tf{35.4} & \tf{36.8} \\
\midrule
\multirow{2}{*}{\tf{ANCE}}
& $s_{q\text{-}d}$ & 43.4 & 43.3& 28.4 & 23.2& 43.2 & 47.1& 7.8 & 11.0& 56.8 & 61.2& 35.9 & 37.1\\
& \shorttitle & \tf{48.4} & \tf{48.9} & \tf{30.5} & \t{24.6} & \tf{46.3} & \tf{50.1} & \tf{8.6} & \tf{11.9} & \tf{65.2} & \tf{68.7}& \tf{39.8} & \tf{40.9} \\
\midrule
\multirow{2}{*}{\tf{TAS-B}}
& $s_{q\text{-}d}$ & 66.4 & 67.6& 39.6 & 31.9& 62.5 & 66.2& 12.1 & 16.5& 75.4 & 77.7& 51.2 & 52.0 \\
& \shorttitle & \tf{69.2} & \tf{71.1} & \tf{40.8} & \tf{32.9} & \tf{64.1} & \tf{67.7} & \tf{12.5} & \tf{16.8} & \tf{81.2} & \tf{82.7} & \tf{53.6} & \tf{54.2} \\
\midrule
\multirow{2}{*}{\tf{GTR}}
& $s_{q\text{-}d}$ & 61.6 & 61.9& 39.9 & 32.3& 58.8 & 62.3& 12.0 & 16.2& 72.6 & 75.6& 49.0 & 49.7 \\
& \shorttitle & \tf{65.5} & \tf{66.8} & \tf{41.0} & \tf{33.5} & \tf{62.0} & \tf{65.3} & \tf{12.2} & \tf{16.7} & \tf{82.4} & \tf{83.4} & \tf{52.6} & \tf{53.1} \\
\bottomrule
\end{tabular}%
}
\caption{Document Retrieval Performance (nDCG@$k$ = 5, 20 in percentage, abbreviated as ND@$k$) on all the queries, including the ones containing single and multiple subqueries. The setup is similar to Table \ref{tab:ir}. comparing the retrieval results based on $s_{q\text{-}d}$ and \shorttitle . \tf{Bold} presents the best performance across the metrics. We can notice that \shorttitle outperform the previous query-document retrieval for each entry.}
\label{tab:ir_single}
\end{table*}

\begin{table}[h]
\centering
\begin{tabular}{lcc}
\hline
Model     & Avg. \#prop in \( r_{q\text{-}d} \prec r_{s\text{-}p} \) & Avg. \#prop in \( r_{q\text{-}d} \succ r_{s\text{-}p} \) \\ \hline
SimCSE    & 9.06 & 6.32 \\
Contriever & 8.25 & 7.24 \\
ANCE      & 8.12 & 8.15 \\
DPR       & 8.54 & 7.88 \\
GTR       & 8.45 & 6.79 \\
TAS-B     & 8.00 & 7.52 \\ \hline
\end{tabular}
\caption{Average number of propositions in two sets of document for different retrievers, i.e., $r_{q\text{-}d} \prec r_{s\text{-}p}$ and $r_{q\text{-}d} \succ r_{s\text{-}p}$. We can notice the average number of propositions in $r_{q\text{-}d} \prec r_{s\text{-}p}$ is more than $r_{q\text{-}d} \succ r_{s\text{-}p}$. This shows that the finer granularity can better deal with the documents with more propositions.}
\label{tab:two_sets}
\end{table}

\section{\shorttitle for Other Domains} \label{appendix:other_domains}

Our work provides a comprehensive analysis of the impact of \shorttitle on scientific text retrieval, considering both the variety of datasets and the use of dense retrievers.
The applicability of \shorttitle to other domains remains an open question.
We explore this by conducting document retrieval experiments on three distinct datasets: ConditionalQA \cite{sun-etal-2022-conditionalqa}, FiQA \cite{maia201818}, and Arguana \cite{wachsmuth-etal-2018-retrieval}, which belong to the domains of law, finance, and argumentation, respectively.
The results are detailed in Table \ref{tab:other_domains}.
We observe that \shorttitle's benefits are considerably more limited, or even negative, outside the scientific context.
This disparity may be attributed to the varying degrees of alignment between the domain-specific characteristics of each field and the training corpus of the dense retrievers.
Or, \ti{propositioner} can not perform well in these domains.
Such findings further underscore the potentially distinct domain-specific nature of scientific document retrieval.

\begin{table*}[h]
\centering
\resizebox{0.95\textwidth}{!}{%
\begin{tabular}{lc|cc|cc|cc|cc}
\toprule
\multirow{2}{*}{\tf{Retriever}} & {\multirow{2}{*}{\tf{Setup}}} & \multicolumn{2}{c|}{\tf{Arguana}} & \multicolumn{2}{c|}{\tf{ConditionalQA}} & \multicolumn{2}{c|}{\tf{FiQA}} & \multicolumn{2}{c}{\tf{Avg.}} \\
 & & ND@5  & ND@20 & ND@5  & ND@20 & ND@5  & ND@20 & ND@5  & ND@20 \\ 
\midrule
\multicolumn{10}{c}{Unsupervised Dense Retrievers} \\
\midrule
\multirow{4}{*}{\tf{SimCSE}}    & $s_{q\text{-}d}$ & 16.4 & 25.9& 52.3 & 58.0& 8.4 & 10.9& 25.7 & 31.6 \\
& $s_{q\text{-}p}$ & 12.5 & 20.9& 53.7 & 59.5& 7.6 & 9.7& 24.6 & 30.0 \\
& $s_{s\text{-}p}$ & 6.3 & 12.3& 42.8 & 50.8& 9.3 & 11.6& 19.5 & 24.9 \\
& \shorttitle      & 12.7 & 22.4& 57.7 & 63.3& 10.6 & 13.8& 27.0 & 33.2 \\

\midrule
\multirow{4}{*}{\tf{Contriever}} & $s_{q\text{-}d}$ & 25.9 & 36.0& 82.5 & 83.9& 25.0 & 29.9& 44.5 & 49.9 \\
                           & $s_{q\text{-}p}$ & 24.8 & 35.9& 81.8 & 83.5& 18.8 & 23.1& 41.8 & 47.5 \\
                           & $s_{s\text{-}p}$ & 24.1 & 34.5& 63.3 & 67.2& 18.6 & 22.9& 35.3 & 41.5 \\
                           & \shorttitle & 28.7 & 39.2& 83.5 & 84.5& 24.7 & 29.8& 45.6 & 51.2 \\

\midrule
\multicolumn{10}{c}{Supervised Dense Retrievers} \\
\midrule
\multirow{4}{*}{\tf{DPR}}       & $s_{q\text{-}d}$ & 9.0 & 16.6& 58.5 & 63.6& 12.0 & 14.6& 26.5 & 31.6 \\
                           & $s_{q\text{-}p}$ & 8.4 & 16.9& 60.1 & 64.7& 8.4 & 10.9& 25.6 & 30.8 \\
                           & $s_{s\text{-}p}$ & 6.1 & 12.2& 34.8 & 41.8& 9.2 & 11.8& 16.7 & 21.9 \\
                           & \shorttitle & 8.2 & 16.3& 59.9 & 65.4& 11.2 & 14.9& 26.4 & 32.2 \\
\midrule
\multirow{4}{*}{\tf{ANCE}}      & $s_{q\text{-}d}$ & 12.0 & 20.5& 64.2 & 68.0& 14.6 & 18.2& 30.3 & 35.6\\
                           & $s_{q\text{-}p}$ & 11.7 & 21.3& 64.0 & 68.2& 8.5 & 10.9& 28.1 & 33.5 \\
                           & $s_{s\text{-}p}$ & 10.1 & 18.6& 41.4 & 48.1& 8.4 & 11.3& 20.0 & 26.0 \\
                           & \shorttitle & 12.4 & 21.8& 66.2 & 69.8& 12.8 & 16.2& 30.5 & 36.0 \\
\midrule
\multirow{4}{*}{\tf{TAS-B}}     & $s_{q\text{-}d}$ & 27.9 & 37.8& 75.3 & 77.9& 26.7 & 31.5& 43.3 & 49.0 \\
                           & $s_{q\text{-}p}$ & 18.8 & 30.5& 76.4 & 78.7& 15.3 & 19.7& 36.8 & 43.0 \\
                           & $s_{s\text{-}p}$ & 12.9 & 20.8& 60.8 & 65.2& 13.9 & 17.8& 29.2 & 34.6 \\
                           & \shorttitle & 22.6 & 33.6& 77.7 & 79.2& 22.8 & 27.9& 41.1 & 46.9 \\

\midrule
\multirow{4}{*}{\tf{GTR}}       & $s_{q\text{-}d}$ & 31.4 & 40.7& 79.8 & 82.3& 34.4 & 39.6& 48.5 & 54.2 \\
                           & $s_{q\text{-}p}$ & 25.6 & 36.9& 80.1 & 82.0& 22.8 & 27.4& 42.8 & 48.8 \\
                           & $s_{s\text{-}p}$ & 20.4 & 30.0& 62.9 & 67.7& 19.6 & 24.2& 34.3 & 40.6 \\
                           & \shorttitle & 29.4 & 39.4& 82.4 & 84.1& 30.8 & 36.1& 47.5 & 53.2 \\
\bottomrule
\end{tabular}%
}
\vspace{-0.07in}
\caption{Comparison between \shorttitle and its components on ConditionalQA, Arguana, and FiQA. We can find that the similarity based on the finer granularity $s_{s\text{-}p}$ and \shorttitle won't bring as many benefits as their performance in the scientific domains, even the degradation.}
\label{tab:other_domains}
\end{table*}

\newblock
\section{Licences of Scientific Artifacts}
\label{appendix:artifacts}
\definecolor{mygray}{gray}{.9}
\begin{table*} 
    \centering
    \resizebox{\textwidth}{!}{
    \begin{tabular}{llll} 
       \toprule
        \textbf{Artifacts/Packages} & \textbf{Citation} & \textbf{Link} & \textbf{License}\\ 
        \rowcolor{mygray} \multicolumn{4}{c}{\textit{Artifacts(datasets/benchmarks). }}\\
        SciFact & \cite{wadden-etal-2020-fact} & \url{https://huggingface.co/datasets/BeIR/scifact} & cc-by-sa-4.0 \\
        SciDocs & \cite{cohan-etal-2020-specter} & \url{https://huggingface.co/datasets/BeIR/scidocs} & cc-by-sa-4.0 \\
        SciQ & \cite{welbl-etal-2017-crowdsourcing} & \url{https://huggingface.co/datasets/bigbio/sciq} & cc-by-nc-3.9 \\
        NFCorpus & \cite{boteva2016full} & \url{https://huggingface.co/datasets/BeIR/nfcorpus} & cc-by-sa-4.0 \\
        \rowcolor{mygray} \multicolumn{4}{c}{\textit{Packages}}\\
        PyTorch & \cite{paszke-etal-2019-pytorch} & \url{https://pytorch.org/} & BSD-3 License\\
        transformers & \cite{wolf2019huggingface} & \url{https://huggingface.co/transformers/v2.11.0/index.html} & Apache License 2.0\\
        numpy & \cite{DBLP:journals/nature/HarrisMWGVCWTBS20} & \url{https://numpy.org/} & BSD License \\
        matplotlib & \cite{hunter2007matplotlib} & \url{https://matplotlib.org/} & BSD compatible License\\
        vllm & \cite{kwon2023efficient} & \url{https://github.com/vllm-project/vllm} & Apache License 2.0 \\
       \rowcolor{mygray} \multicolumn{4}{c}{\textit{Models}}\\
        LLaMA-3 & \cite{DBLP:journals/corr/abs-2307-09288} & \url{https://huggingface.co/meta-llama/Meta-Llama-3-8B-Instruct} & \href{https://ai.meta.com/llama/license/}{LICENSE}\\
        SimCSE & \cite{gao-etal-2021-simcse} & \url{https://huggingface.co/princeton-nlp/unsup-simcse-bert-base-uncased} & MIT license \\
        Contriever & \cite{izacard2022unsupervised} & \url{https://huggingface.co/facebook/contriever} & \href{https://github.com/facebookresearch/contriever?tab=License-1-ov-file\#readme}{License} \\
        DPR & \cite{karpukhin-etal-2020-dense} & \url{https://huggingface.co/facebook/dpr-ctx_encoder-multiset-base} & cc-by-nc-4.0 \\
        ANCE & \cite{xiong2021approximate} & \url{https://huggingface.co/castorini/ance-dpr-context-multi} & MIT license\\
        TAS-B & \cite{hofstatter2021efficiently} & \url{https://huggingface.co/sentence-transformers/msmarco-distilbert-base-tas-b} & Apache License 2.0\\
        GTR & \cite{ni-etal-2022-large} & \url{https://huggingface.co/sentence-transformers/gtr-t5-base} & Apache License 2.0\\
      \bottomrule
    \end{tabular}}
    \caption{Details of datasets, major packages, and existing models we use. The datasets we reconstructed or revised and the code/software we provide are under the MIT License.}
    \label{tab:tools}
\end{table*}

\end{document}